\documentclass{article}
\usepackage{amssymb,graphicx,cancel}
\usepackage{jheppub}
\usepackage{slashed,color}
\usepackage[table]{xcolor}
\usepackage[utf8]{inputenc}
\usepackage{booktabs,multirow,amssymb}
\usepackage[export]{adjustbox}
\usepackage{floatrow,blindtext,amsmath,booktabs,multicol}
\usepackage{lipsum,marvosym,subfig,array}
\usepackage[normalem]{ulem}
\usepackage{tabularx,cleveref}

\def\gsim{\mathrel{\raise.3ex\hbox{$>$\kern-.75em\lower1ex\hbox{$\sim$}}}}
 \DeclareMathOperator{\ev}{eV}                    \DeclareMathOperator{\few}{few} 
          \newcommand{\cP}{{ \cal P}}   
\newcommand{\ep}{\epsilon}  
    
\newcommand{\pL}{\left(} \newcommand{\pR}{\right)} \newcommand{\bL}{\left[} \newcommand{\bR}{\right]}    
\newcommand{\beq}{\begin{equation}} \newcommand{\eeq}{\end{equation}}
\newcommand{\bea}{\begin{eqnarray}} \newcommand{\eea}{\end{eqnarray}}
\newcommand{\alg}[1]{\begin{align} \begin{split} #1 \end{split}  \end{align}}

\newcommand{\tenx}[1]{\times 10^{#1}}
\newcommand{\Eq}[1]{Eq.~(\ref{#1})}  \newcommand{\Eqm}[2]{Eqs.~(\ref{#1}) through (\ref{#2})}
\newcommand{\Sec}[1]{Sec.~\ref{#1}}  
\newcommand{\Fig}[1]{Fig.~\ref{#1}} 

\newcommand{\App}[1]{App.~\ref{#1}}

\newcommand{\eg}{\emph{e.g.~}}
\newcommand{\ie}{\emph{i.e.~}}

\title{Dark Photon Dark Matter in the Presence of Inhomogeneous Structure }
\author{Samuel J.~Witte${}^\dagger$,}
\emailAdd{Samuel.Witte@ific.uv.es}
\affiliation{${}^\dagger$Instituto de Fisica Corpuscular (IFIC), CSIC-Universitat de Valencia, Spain}
\author{Salvador Rosauro-Alcaraz${}^\star$,}
\emailAdd{Salvador.Rosauro@uam.es}
\affiliation{${}^\star$ Departamento de F\'{i}sica T\'{e}orica and Instituto de F\'{i}sica T\'{e}orica, IFT-UAM/CSIC,
Universidad Aut\'{o}noma de Madrid, Cantoblanco, 28049, Madrid, Spain}
\author{Samuel D.~McDermott${}^\ddagger$,}
\emailAdd{SamMcD00@fnal.gov}
\affiliation{${}^\ddagger$Theoretical Astrophysics Group, Fermi National Accelerator Laboratory, Batavia, IL, USA}
\author{Vivian Poulin${}^\Box$}
\emailAdd{Vivian.Poulin@umontpellier.fr}
\affiliation{${}^\Box$LUPM, CNRS \& Universit\'e de Montpellier, F-34095 Montpellier, France}

\begin{document}
\preprint{\begin{flushright}FERMILAB-PUB-20-121-T \\ IFT-UAM/CSIC-20-47 \\ FTUAM-20-7 \\ LUPM:20-016\end{flushright}}
\abstract{ Dark photon dark matter will resonantly convert into visible photons when the dark photon mass is equal to the plasma frequency of the ambient medium. In cosmological contexts, this transition leads to an extremely efficient, albeit short-lived, heating of the surrounding gas. Existing work in this field has been predominantly focused on understanding the implications of these resonant transitions in the limit that the plasma frequency of the Universe can be treated as being perfectly homogeneous, \ie neglecting inhomogeneities in the electron number density. In this work we focus on the implications of heating from dark photon dark matter in the presence of inhomogeneous structure (which is particularly relevant for dark photons with masses in the range $10^{-15} \,  {\rm eV} \, \lesssim m_{A^\prime} \lesssim 10^{-12}$ eV), emphasizing both  the importance of inhomogeneous energy injection, as well as the sensitivity of cosmological observations to the inhomogeneities themselves. More specifically, we derive modified constraints on dark photon dark matter from the Ly-$\alpha$ forest, and show that the presence of inhomogeneities allows one to extend constraints to masses outside of the range that would be obtainable in the homogeneous limit, while only slightly relaxing their strength.  We then project sensitivity for near-future cosmological surveys that are hoping to measure the 21cm transition in neutral hydrogen prior to reionization, and demonstrate that these experiments will be extremely useful in improving sensitivity to masses near $\sim 10^{-14}$ eV, potentially by several orders of magnitude. Finally, we discuss implications for reionization, early star formation, and late-time $y$-type spectral distortions, and show that probes which are inherently sensitive to the inhomogeneous state of the Universe could resolve signatures unique to the light dark photon dark matter scenario, and thus offer a fantastic potential for a positive detection. }

\maketitle

\setcounter{page}{2}

\section{Introduction}

 There is overwhelming evidence for the existence of dark matter  on a wide variety of astrophysical scales based solely on its gravitational influence. Despite extensive theoretical and experimental efforts over the past four decades however, non-gravitational signatures of dark matter, should they exist, have yet to be robustly identified. Therefore the identification of the true nature of dark matter, which is among the strongest pieces of evidence for the existence of physics beyond the Standard Model (SM), is still lacking. In recent years, there has been an increasing interest in the dark matter community to remove theoretical prejudice in how and where we search for dark matter;  the mentality of `leaving no stone unturned' has carried with it a renewed interested in the low-energy / high-intensity frontier. Dark matter with a sub-eV mass must be non-thermally produced in order to be sufficiently cold, and bosonic in order to fit the abundance observed inside low-mass gravitationally bound objects. For many years, the leading candidate that  satisfies these conditions has been the axion (see \eg Refs.~\cite{Marsh:2015xka,Irastorza:2018dyq} for reviews on axions). An alternative possibility which has gained increasing interest is that dark matter is comprised of light vector bosons~\cite{Jaeckel:2008fi,Pospelov:2008jk,Redondo:2008ec,Mirizzi:2009iz,Nelson:2011sf,Arias:2012az,Graham:2015rva,Dubovsky:2015cca,Agrawal:2018vin, Dror:2018pdh, Co:2018lka, Bastero-Gil:2018uel, Bhoonah:2018gjb, Kovetz:2018zes, Long:2019lwl, AlonsoAlvarez:2019cgw, Nakayama:2019rhg, Wadekar:2019xnf, McDermott:2019lch, Caputo:2020bdy}. From the perspective of particle physics, a particularly simple dark matter candidate that has a non-trivial coupling to the Standard Model is a dark photon $A^\prime_\mu$, which kinetically mixes with the SM photon via the renormalizable operator $\ep \, F^{\mu \nu} \, F'_{\mu \nu}\, / \, 2$ \cite{Holdom:1985ag}.

Historically, one of the concerns that has limited the appeal of this candidate was its production mechanism. Unlike the axion, the misalignment mechanism cannot be used to efficiently produce light vector bosons~\cite{Nelson:2011sf} because the norm of the vector field, and thus the energy density, is efficiently diluted in the early Universe~\cite{Arias:2012az,Graham:2015rva,AlonsoAlvarez:2019cgw}. This can be avoided by introducing a non-minimal coupling to the Ricci scalar, but this fix comes at the cost of introducing instabilities in the longitudinal mode of the dark photon~\cite{Himmetoglu:2008zp,Himmetoglu:2009qi,Karciauskas:2010as}. Field excitations induced during inflation were shown to be capable of producing the correct abundance; however, this mechanism over-predicts primordial gravitational waves if the mass of the dark photon $m_{A^\prime} \gtrsim \mu{\rm eV}$~\cite{Graham:2015rva}. More recently, a number of novel production mechanisms were shown to be efficient in producing dark photons of sub-eV mass down to the fuzzy dark matter scale $\sim 10^{-21}$ eV. These proposals can be broken into two categories: the first of these exploits a tachyonic instability that arises when the dark photon couples to a decaying scalar (\eg the inflation during reheating)~\cite{Agrawal:2018vin, Dror:2018pdh, Co:2018lka, Bastero-Gil:2018uel}, and the second relies on the fact that cosmic strings may preferentially radiate dark photons~\cite{Long:2019lwl}. These proposals
exhibit a wide range of possible early-universe phenomenology, leading to renewed interest in the late-universe behavior of light vector boson dark matter.

Phenomenologically, light dark photons are unique because the efficiency with which they interact with SM particles depends strongly on the plasma frequency of the surrounding medium, $\omega_p$, a quantity which scans $\sim 15$ orders of magnitude between big bang nucleosynthesis (BBN) and today. As a result of the kinetic mixing between the dark and visible photons, interactions are resonantly enhanced when their mass ordering changes~\cite{Redondo:2008ec}.

The vast majority of  previous work on the cosmological implications of this resonance have effectively assumed that the plasma frequency of the Universe is homogeneous~\cite{Arias:2012az, McDermott:2019lch} -- the simplifying assumption is that $\omega_p$, and thus the {\it time} of resonance, is a function exclusively of {\it redshift}. While this is likely to be approximately valid at redshifts $z \gtrsim 100$, when almost all density perturbations remain linear, the formation of structure at lower redshift strongly violates this assumption and can have wide-reaching implications for experiments with sensitivity to ultra-light dark photon dark matter. One immediate implication of properly accounting for the presence of inhomogeneous structure is that resonant constraints will extend over a wider range of masses -- this is simply because, at any instant in time, the plasma frequency in voids (halos) can be significantly smaller (larger) than the mean plasma frequency of the Universe, thus allowing for a wider range of masses to undergo resonant conversion. This is of particular importance in the case of voids, as bounds derived in the homogeneous limit typically have an abrupt edge at low masses \cite{Arias:2012az, McDermott:2019lch}, and cosmic voids offer a cosmological laboratory by which these bounds can be smoothly extended \cite{Caputo:2020bdy}. This also has interesting implications for the recent claim by the EDGES collaboration of the observation of an anomalous absorption dip of the 21cm line at high redshift by neutral hydrogen \cite{Bowman:2018yin,Pospelov:2018kdh},  as shown in Refs.~\cite{Pospelov:2018kdh, Bondarenko:2020moh, Garcia:2020qrp} when the dark photons are not cold.

In this work, we investigate the extent to which the presence of inhomogeneities modifies the energy injection arising from both resonant and non-resonant conversion of dark photon dark matter. We illustrate the potential importance of structure by generalizing constraints derived from excess heating of the intergalactic medium (IGM) prior to and during the epoch of helium reionization (obtained using observations of the Ly-$\alpha$ forest) to account for inhomogeneities, showing that the presence of structure serves to broaden constraints over a wider range of dark photon masses. We then show that future radio telescopes aiming to measure the  21cm differential brightness temperature during the Cosmic Dawn may be able to improve upon cosmological constraints by a few orders of magnitude for dark photon masses $m_{A^\prime} \sim 10^{-14}$ eV. Finally, we discuss implications of inhomogeneous heating during the epoch of reionization, the impact of dark photon heating on the formation of the first stars, the extent to which constraints derived using the CMB in the homogeneous limit should be consider robust, and late-time spectral distortions induced from an excess heating of the IGM. This work provides novel insights into interesting and unique cosmological signatures that can be used to constrain or confirm the existence of dark photon dark matter.

\section{Dark Photon Conversion in the Presence of Inhomogeneities}\label{sec:dp_overv}
We first begin by reviewing energy injection from dark photon dark matter under the simplifying assumption that the Universe can be treated as homogeneous~\cite{Arias:2012az,McDermott:2019lch}, and then generalize this formalism in the following subsection to account for inhomogeneous structure.

\subsection{The Homogeneous Universe}

The differential energy density per unit redshift introduced to the SM photon bath in a homogeneous (`homo') Universe by a dark photon with mass $m_{A'}$ and mixing $\epsilon$ is
\begin{equation} \label{hom-drhodz-NR}
	\frac{d\rho_{A' \to \gamma}^{\rm homo}}{dz}(z) = \rho_{\rm CDM}^{\rm homo}(z) \, \frac{d}{dz} P_{A^\prime \rightarrow \gamma}^{\rm homo}(z) ,
\end{equation}
where $dP_{A^\prime \rightarrow \gamma}(z)/dz$ is the probability of conversion and we have assumed (as we will throughout this work) that the dark photon accounts for the entirety of dark matter. For non-adiabatic conversion of a dark photon, which is the scenario of interest, the differential conversion probability per unit {\it time} (assuming that the probability of conversion is much less than unity) is given by the Landau-Zener formula~\cite{Zener:1932ws}, which in the limit of small conversion probability is given by \cite{Mirizzi:2009iz}
\begin{equation} \label{hom-dPdz-NR}
	\frac{d}{dt} P_{A^\prime \rightarrow \gamma}^{\rm homo}(z) \simeq \frac{\pi \epsilon^2 m_{A^\prime}^2}{ \omega } \, \left|  \frac{d\ln \big[ \omega_p^{\rm homo} \big]^2}{dt}\right|^{-1} \delta(t-t_{\rm res}^{\rm homo}) \, .
\end{equation}
In \Eq{hom-dPdz-NR}, $\omega_p^{\rm homo}$ is the plasma frequency and $t_{\rm res}$ is the time at which the resonant condition is met, which is determined by the condition $m_{A^\prime} \simeq \omega_p$\footnote{More generally, the resonant conversion involves equating the mass with the real part of photon self-energy (see \eg Refs.~\cite{An:2013yfc,Chang:2016ntp}). The limit obtained here is valid only for a non-relativistic and non-degenerate plasma. }. At leading order, the plasma frequency is given by
\begin{equation}
	\omega_p(\vec{x}, z) = \sqrt{\frac{4\pi\alpha \, n_e(\vec{x}, z)}{m_e}} \, ,
\end{equation} 
which in the context of a homogeneous Universe reduces to $\omega_p^{\rm homo}(z) = \overline{\omega_p(\vec{x}, z)}$, with $\overline{\omega_p(\vec{x}, z)}$ representing the spatial average. Since $\omega_p$ in a homogeneous Universe inherits time-dependence solely via the effect of the expansion rate on the electron number density, we may further simplify \Eq{hom-dPdz-NR} by writing
\begin{equation}\label{eq:prob2}
	\frac{d}{dz} P_{A^\prime \rightarrow \gamma}^{\rm homo}(z) \simeq \frac{\pi \epsilon^2 m_{A^\prime}}{ 3 H(z)} \delta(z-z_{\rm res}^{\rm homo}) \, .
\end{equation} 
In \Eq{eq:prob2} we have introduced $z_{\rm res}^{\rm homo}$, the redshift at which a dark photon of a given mass undergoes resonance. This is given by solving
\begin{equation}
	n_e^{\rm homo}(z_{\rm res}^{\rm homo}) = \frac{m_e \, m_{A^\prime}^2}{4\pi\alpha} \, ,
\end{equation}
where the homogeneous electron number density is given by
\begin{equation}
	n_e^{\rm homo}(z_{\rm res}^{\rm homo}) = x_e^{\rm homo}(z_{\rm res}^{\rm homo}) (1 - Y_p / 2) \eta \frac{2 \zeta(3)}{\pi^2} T_0^3 (1+z_{\rm res}^{\rm homo})^3 \, .
\end{equation}
Here, $x_e$ is the free electron fraction, $Y_p \simeq 0.245$ is the primordial helium abundance \cite{Aver:2015iza, Pitrou:2018cgg}, $\eta \simeq 6.1\tenx{-10}$ is the baryon-to-photon ratio \cite{Fields:2019pfx}, and $T_0 \simeq 2.2755$ K the CMB temperature today \cite{Fixsen:2009ug}.

Anticipating the scenario of interest below, it is trivial to extend \Eqm{hom-drhodz-NR}{eq:prob2} to consider the energy deposited per baryon as a function of redshift:
\begin{equation} \label{hom-drhoPerBdz-NR}
	\frac{d}{dz} \frac{\rho_{A' \to \gamma}^{\rm homo}}{n_b^{\rm homo}}(z) \simeq \frac{\pi \epsilon^2 m_{A'}}{ 3 H(z) } \frac{\rho_{\rm CDM}^{\rm homo}}{n_b^{\rm homo}}  \delta(z-z_{\rm res}^{\rm homo})  \, ,
\end{equation}
where $n_b^{\rm homo}$ is the baryon number density in a homogeneous Universe. \Eq{hom-drhoPerBdz-NR} can be integrated over a redshift range of interest to obtain the homogeneous result for the specific energy injected per baryon at a given cosmological epoch, which was used in Ref.~\cite{McDermott:2019lch} to constrain exotic energy injection by dark photons using observations of the Ly-$\alpha$ forest.

As described in  Ref.~\cite{Dubovsky:2015cca,McDermott:2019lch}, even if the dark photon mass is small compared to the local plasma frequency, dark photons can deposit energy into the medium via an inverse bremsstrahlung process, with the rate of energy deposition being suppressed by the ratio $(m_{A^\prime}/\omega_p)^2$. The probability of absorption from this process in a homogeneous medium can be written as
\begin{equation} \label{Q-dp}
P_{A' \to \gamma}^{\rm IB, homo}  \simeq  \frac{ \ep^2 \nu^{\rm homo}}{2(1+z) \, H(z)} \bL \frac{m_{A'}^2}{\omega_p^{\rm homo}(z)^2} \bR^{{\rm sign}[\omega_p^{\rm homo}(z)-m_{A'}]} \, ,
\end{equation}
with the frequency of electron-ion collisions $\nu^{\rm homo}$ given by
\begin{equation}
\nu^{\rm homo} = \frac{4\, \sqrt{2\pi} \, \alpha_{\rm EM}^2 \, n_e^{\rm homo}}{3\, \sqrt{m_e \, (T_k^{\rm homo})^3 }}\,  \log \pL \sqrt{ \frac{4 \pi \, (T_k^{\rm homo})^3}{\alpha_{\rm EM}^3 \, n_e^{\rm homo}} }\pR \, .
\end{equation}
It is straightforward to generalize \Eq{hom-drhodz-NR} and \Eq{hom-drhoPerBdz-NR} to the case of energy injection and specific energy injection for the inverse bremsstrahlung process in a homogeneous Universe. This process is much less efficient than that of resonant conversion, and is thus of interest only for dark photons with extremely low masses (\ie below the minimum plasma frequency of the Universe).

\subsection{The Inhomogeneous Universe} \label{sec:inhomoU}

We now extend this formalism to the case of an inhomogeneous Universe.  The presence of inhomogeneities broadens the resonance such that a dark photon of a given mass will be able to undergo resonant conversion over a redshift interval rather than at a fixed value of $z$; alternatively, one can understand that the effect of inhomogeneities is to  induce a spatial dependence of $\omega_p$ such that dark photons over a broad range of masses will be capable of undergoing resonant conversion at any given redshift. Introducing the overdensity $\Delta_b \equiv \rho_b / \bar{\rho}_b$, we may decompose the electron number density as $n_e(\vec{x}, z) = n^{\rm homo}_e(z) \times \Delta_b$. Thus, the resonance condition $m_{A^\prime} \simeq \omega_p(\vec{x}, z) = \omega_p^{\rm homo}(z) \times \sqrt{\Delta_b}$ and therefore also the conversion probability $P_{A' \to \gamma}$ will depend on the local value of $\Delta_b$. In particular, resonance will be occur in voids at earlier times ($z > z_{\rm res}^{\rm homo}$) and in halos at later times ($z < z_{\rm res}^{\rm homo}$).

More quantitatively, we generalize the energy injection per unit redshift from \Eq{hom-drhodz-NR} by introducing a probability density function characterizing the baryonic density perturbations at a given redshift $P_\Delta(z, \Delta_b)$.  This allows us to write
\alg{ \label{INhom-drhodz-NR}
	\frac{d\rho_{A' \to \gamma}}{dz}(z)& =  \int \, d\Delta_b \, P_\Delta(z, \Delta_b) \,  \rho_{\rm CDM}(z) \, \frac{d}{dz} P_{A^\prime \rightarrow \gamma}(z, \Delta_b) 
	\\ & = \rho^{\rm homo}_{\rm CDM}(z) \int \, d\Delta_b \,   \Delta_b  \, P_\Delta(z, \Delta_b)  \, \frac{d}{dz} \, P_{A^\prime \rightarrow \gamma}(z, \Delta_b)  \, ,
}
where in the second line we have assumed $\Delta_{\rm CDM} = \Delta_b$.
The differential probability of dark photons converting per unit redshift in the inhomogeneous case differs from \Eq{hom-dPdz-NR} in several ways: {\emph{(i)} } the time at which dark photons undergo resonance now depends on the local overdensity, $t_{\rm res}^{\rm homo} \rightarrow t_{\rm res}(\Delta_b)$, and {\emph{(ii)}} the $d \ln \omega_p^2/dt$ term has additional contributions from the time-dependent evolution of over-densities as well as the relative motion of the dark photon and baryon fluids. 
Since we focus on the case of non-relativistic dark photon conversion after recombination\footnote{For instance, this allows us to ignore the bulk relative velocity  $v\sim 30(1+z/1000)$km/s at $z<1000$.}, the typical distance travelled by the dark matter per Hubble time is extremely small. Furthermore, in the presence of inhomogeneities, one can generalize the time derivative term in \Eq{hom-dPdz-NR} to 
\begin{equation}
	\frac{d\ln\omega_p^2}{dt} = \frac{d\ln\omega_p^2}{dz}\frac{dz}{dt} +   \frac{d\ln\omega_p^2}{dx}\frac{dx}{dt} \simeq \frac{dz}{dt}  \frac{d\ln([\omega_p^{\rm homo}]^2 \, \times \Delta_b)}{dz} \simeq H(z) (1+z) \left(\frac{3}{1+z} +  \frac{d\ln\Delta_b}{dz}\right) \, ,
\end{equation}
where in the last step we have assumed that the time-dependence of $x_e$ is small. Accounting for the fact that an overdensity $\delta_b \equiv (\Delta_b - 1)$ grows proportionally to $(1+z)^{-1}$, we find 
\begin{equation}\label{eq:delBcont}
	\frac{d\ln\Delta_b}{dz} = \frac{1}{1+\delta_b / (1+z)} \, \frac{-\delta_b}{(1+z)^2}\,.
\end{equation}
Thus, the contribution from \Eq{eq:delBcont} is never more than one-third as large as the part coming from the $d\ln[\omega_p^{\rm homo}]^2/dz$ term, and is typically much smaller. In what follows, we choose to neglect these small corrections to $d\ln[\omega_p^{\rm homo}]^2/dz$.

The differential probability of conversion for a dark photon with mass $m_{A^\prime}$ per unit redshift is therefore
\begin{equation}\label{eq:inhomoprob}
	\frac{d}{dz} P_{A^\prime \rightarrow \gamma}(z, \Delta_b) \simeq \frac{\pi \epsilon^2 m_{A^\prime}}{ 3 H(z)} \delta\left(z-z_{\rm res}(\Delta_b, m_{A^\prime})\right).
\end{equation}
In \Eq{eq:inhomoprob}, we have introduced the notation $z_{\rm res}(\Delta_b, m_{A^\prime})$ to emphasize that, for a fixed dark photon mass, the resonant transition occurs at a redshift determined by the local baryonic overdensity. An essential observation of our work is that {\it for a given  $\Delta_b$} there will exist a {\it unique redshift $z_{\rm res}(\Delta_b, m_{A^\prime})$} that allows the resonant conversion to take place (modulo the effect of reionization, which we discuss below), since, regardless of the initial magnitude of the inhomogeneity, the local physical baryon density decreases {\it monotonically} with redshift (until becoming non-linear). The value of $z_{\rm res}$ for a particular overdensity and dark photon mass is given by solving
\begin{equation}\label{eq:ZRES}
\frac{4\,\pi\,\alpha \, n_e(z_{\rm res}, \vec{x})}{m_e} = m_{A^\prime}^2  \, ,
\end{equation}
where the electron number density can be expressed as 
\begin{equation}
	n_e(z, \vec{x}) =x_e(z, \vec{x})\, \frac{\rho_b^{\rm homo}(z) }{m_p} \, \Delta_b(\vec{x}) \, (1-Y_p/2) \, .
\end{equation}
We will assume throughout this work that the free electron fraction is homogeneous $x_e(z,\vec{x}) \simeq \bar{x}_e(z)$; we discuss the implications of inhomogeneous reionization later in \Sec{sec:reion}. Notice that, for a fixed dark photon mass, one can equivalently conceptualize this as having a resonant overdensity $\Delta_{\rm res}$ which is a function of redshift.
The differential energy injection per unit volume per unit redshift can then be directly determined with \Eq{INhom-drhodz-NR}; explicitly,
\begin{equation} \label{drhodzddb}
	\frac{d \rho_{A' \to \gamma}}{dz } (z) \simeq \rho^{\rm homo}_{\rm CDM}(z)  \frac{\pi \epsilon^2 m_{A^\prime}}{3 H(z)}	 \, \int \, d\Delta_b \, \,  P_\Delta(z, \Delta_b) \,  \Delta_b  \, \delta(z - z_{\rm res}(\Delta_b, m_{A^\prime})) \, .
\end{equation}
This provides the generalization of the homogeneous energy injection rate.

Integrating \Eq{drhodzddb} over a redshift interval $[z_{\rm low}, z_{\rm high}]$ gives the energy injected per unit volume,
\begin{equation} \label{rhozdb}
\left.\Delta\rho \right|_{z_{\rm low}}^{z_{\rm high}} = \frac{\pi \epsilon^2 m_{A^\prime}}{3} \, \int \, d\Delta_b \, \Delta_b \, \frac{\rho_{\rm CDM}^{\rm homo}\left(z_{\rm res} \right)}{H(z_{\rm res})}\, P_\Delta(z_{\rm res}, \Delta_b)\, \Theta(z_{\rm res} - z_{\rm low}) \, \Theta(z_{\rm high} - z_{\rm res}) \, , 
\end{equation} 
where $\Theta$ is the Heaviside step function, and we have dropped the explicit dependence of $z_{\rm res}$ on $\Delta_b$ and $m_{A'}$ for simplicity. Analogously, one can write the specific energy injected $\varepsilon_{\rm inj}$, defined as the energy injected per unit volume per unit baryon, as
\begin{equation}
\label{rhozlzh}
\left.\varepsilon_{\rm inj}\right|_{z_{\rm low}}^{z_{\rm high}} \, = \, \frac{\pi \epsilon^2 m_{A^\prime}}{3} \, \int \, d\Delta_b \, \frac{\rho_{\rm CDM}^{\rm homo}(z_{\rm res})}{H(z_{\rm res})\,  n_b^{\rm homo}(z_{\rm res} )}\, P_\Delta(z_{\rm res}, \Delta_b) \Theta(z_{\rm res} - z_{\rm low}) \Theta(z_{\rm high} - z_{\rm res}) \, .
\end{equation}
The explicit dependence on the overdensity parameter has dropped out due to the cancellation with the factor of $\Delta_b$ from the baryon number density. The homogeneous limit can be straightforwardly recovered by taking $P_\Delta(z_{\rm res}, \Delta_b) \to \delta(\Delta_b-1)$. 

It is also possible to generalize the expressions for non-resonant absorption of dark photons via inverse bremsstrahlung in \Eq{Q-dp}. Explicitly, the energy deposition rate is
\begin{equation} \label{Q-dpIN}
	\frac{d\rho_{A' \to \gamma}^{\rm (IB)}}{dz}(z) \simeq {\rho}^{\rm homo}_{\rm CDM}(z)  \int \, d\Delta_b \, \Delta_b \, P_\Delta(z, \Delta_b) \, \frac{ \ep^2 \, \nu(\Delta_b, z)}{2(1+z) \, H(z)} \bL \frac{m_{A'}^2}{\omega_p^{\rm homo}(z)^2 \, \Delta_b} \bR^{{\rm sign}[\omega_p^{\rm homo}(z) \, \sqrt{\Delta_b} -m_{A'}]} \, ,
\end{equation}
with the frequency of electron-ion collisions $\nu$ given by
\begin{equation}
\nu(z, \Delta_b) = \frac{4\, \sqrt{2\pi} \, \alpha_{\rm EM}^2 \, n_e^{\rm homo}(z) \, \Delta_b}{3\, \sqrt{m_e \, T_k(\Delta_b, z)^3 }}\,  \log \pL \sqrt{ \frac{4 \pi \, T_k(\Delta_b, z)^3}{\alpha_{\rm EM}^3 \, n_e^{\rm homo}(z) \Delta_b} }\pR \, .
\end{equation}
Here, we have included an explicit dependence of the matter temperature $T_k$ on the overdensity. Generically one expects the temperature to 
obey $T_k \propto \Delta_b^{\beta}$; when only adiabatic cooling is relevant, the solution is $\beta = 2/3$. As in the case of resonant conversion, the rates of energy injection and specific energy injection are 
\begin{equation}
\Delta  \rho_{A' \to \gamma} = \int \, dz \, \frac{\epsilon^2 \rho_{\rm CDM}^{\rm homo}(z) }{2 (1+z) \, H(z)}\,\int \, d\Delta_b \, \Delta_b \, P_\Delta(z, \Delta_b) \,  \nu(\Delta_b, z) \bL \frac{m_{A'}^2}{\omega_p^{\rm homo}(z)^2 \, \Delta_b} \bR^{{\rm sign}[\omega_p^{\rm homo}(z) \, \sqrt{\Delta_b} -m_{A'}]} \, 
\end{equation} 
and 
\begin{equation}\label{eq:inhomIB}
\varepsilon_{\rm inj} = \int \, dz \, \frac{\epsilon^2 \rho_{\rm CDM}^{\rm homo}(z) }{2 (1+z) \, H(z) \, n_b^{\rm homo}(z)}\,\int \, d\Delta_b \, P(\Delta_b, z) \,  \nu(\Delta_b, z) \bL \frac{m_{A'}^2}{\omega_p^{\rm homo}(z)^2 \, \Delta_b} \bR^{{\rm sign}[\omega_p^{\rm homo}(z) \, \sqrt{\Delta_b} -m_{A'}]} \, , 
\end{equation} 
respectively.

The probability distribution function (PDF) characterizing the baryon overdensity is the final ingredient necessary to describe the energy injection from dark photons. The dark matter density field is known to approximately follow a log-normal distribution~\cite{Coles:1991if,Bernardeau:1994aq,Kayo:2001gu}. We will assume here that on scales sufficiently larger than the Jeans scale the baryons track the underlying dark matter density distribution. Thus, for the sake of being concrete, we assume the baryon overdensity PDF is given by 
\begin{equation}
	P_\Delta(z, \Delta_b) = \frac{1}{\sqrt{2\pi \log(1+\sigma^2)}} \frac{1}{\Delta_b } \, e^{-\frac{\log\left( \Delta_b \sqrt{1+\sigma^2}\right)^2}{2 \log(1+\sigma^2)}} \, ,
\end{equation}
where $\sigma^2$ is the variance of the density field. In general, deviations from this simple parameterization are expected, but were found in Ref.~\cite{Caputo:2020bdy} to be minimal over the range $10^{-2} \leq \Delta_b \leq 10^2$  \footnote{Various types of overdensity PDFs were considered in \cite{Caputo:2020bdy}, including: a log-normal distribution, a Gaussian distribution, a distribution extracted from a hydrodynamic simulations~\cite{Nelson:2018uso,McAlpine:2015tma,McCarthy:2016mry,Genel:2014lma,Foreman:2019ahr,vanDaalen:2019pst}, and an analytic approximation based on spherical collapse~\cite{Ivanov:2018lcg,Valageas:2001zr,Valageas:2001td}. }. Thus, we restrict our attention to this range of over-densities in this work, which in turn translates into a conservative  result.
We compute the mass variance $\sigma^2(R,z)$ by convolving the non-linear matter power spectrum $\cP(k,z)$, obtained from  the code {\tt class}~\cite{Blas:2011rf} with the {\tt Halofit}~\cite{Takahashi:2012em} prescription, with a window function $W(kR)$  that smooths the distribution on the scale $R$,
\begin{equation} \label{smoothing-scale}
	\sigma^2(R,z) = \int \frac{dk}{k} \, \frac{k^3 \cP(k,z)}{2\pi^2} \, W^2(kR) \,.
\end{equation}
Since we assume in our formalism that the baryon distribution traces that of the dark matter, we adopt the smallest smoothing scale for which this assumption is expected to be valid: this is the Jeans scale, which is given by~\cite{mo2010galaxy}
\begin{equation}\label{eq:Jeans}
	\lambda_{\rm Jeans}(z, \Delta_b, T_k) = \frac{2\pi}{k_J} = \frac{\pi (1+z)}{H(z)} \sqrt{\frac{8}{3}} \, c_s(z, T_k, \Delta_b) \,,
\end{equation}
where $c_s$ is the speed of sound which in general depends on redshift, baryon temperature, and over-density. In the following, we take $c_s = \sqrt{\gamma k_B T_k / m_p}$, with adiabatic index $\gamma = 5 / 3\mu$, and $\mu$ the mean molecular weight~\cite{mo2010galaxy}, where the gas temperature contains an implicit dependence on redshift and over-density.
On scales smaller than $\lambda_{\rm Jeans}$ the baryon distribution is expected to differ significantly from that of the dark matter owing to the presence of pressure forces, and thus the adopted power spectrum will become inaccurate. Taking slightly larger smoothing scales typically has a minimal impact on the energy injection history  (see \eg Ref.~\cite{Caputo:2020bdy}), while very large smoothing scales reproduce the homogeneous limit. We have also verified explicitly using {\tt class} that differences in the linear baryon spectrum from the dark matter power spectrum at $z \lesssim 50$ are small on the scales of interest, and at larger redshifts the result tends toward the homogeneous result. For these reasons, we take $R=\lambda_{\rm Jeans}(z, \Delta_b, T_k)$ in \Eq{smoothing-scale}.

\subsection{Impact of Inhomogeneities on Resonance}

\begin{figure*}[t]
	\includegraphics[width=0.6\textwidth]{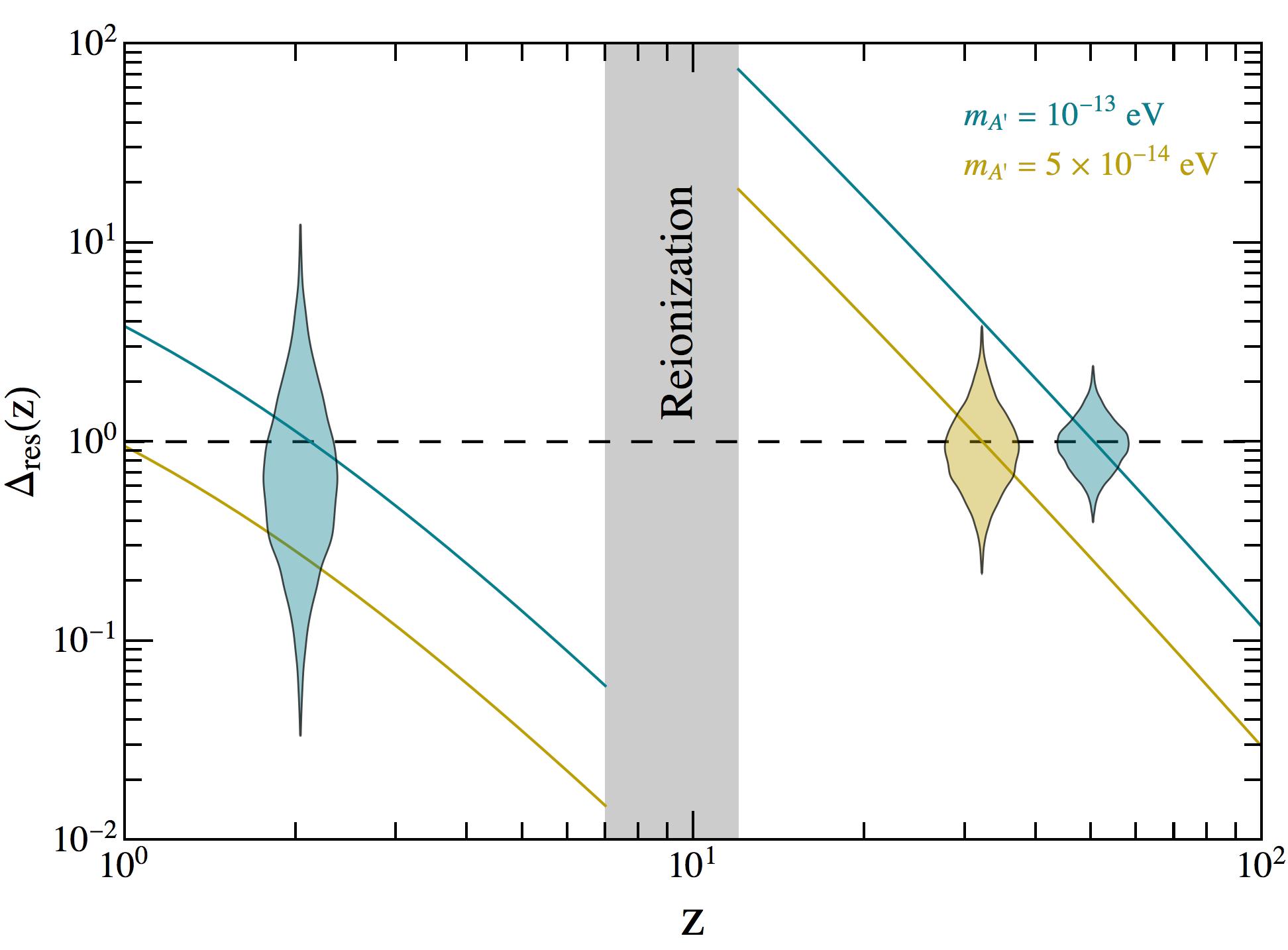}
	\caption{ \label{fig:den_violin} Resonant over-density $\Delta_{\rm res}(z)$ for various dark photon masses. Redshifts $z \in [7, 12] $ have been removed to avoid complications associated with reionization (see \Sec{sec:reion} for further discussion). Normalized log-normal probability distribution functions $P_\Delta$ evaulated at the redshift(s) $z_{\rm res}(\Delta_b = 1)$ are shown for comparison.   }
\end{figure*}

\begin{figure*}[t]
	\includegraphics[width=0.6\textwidth]{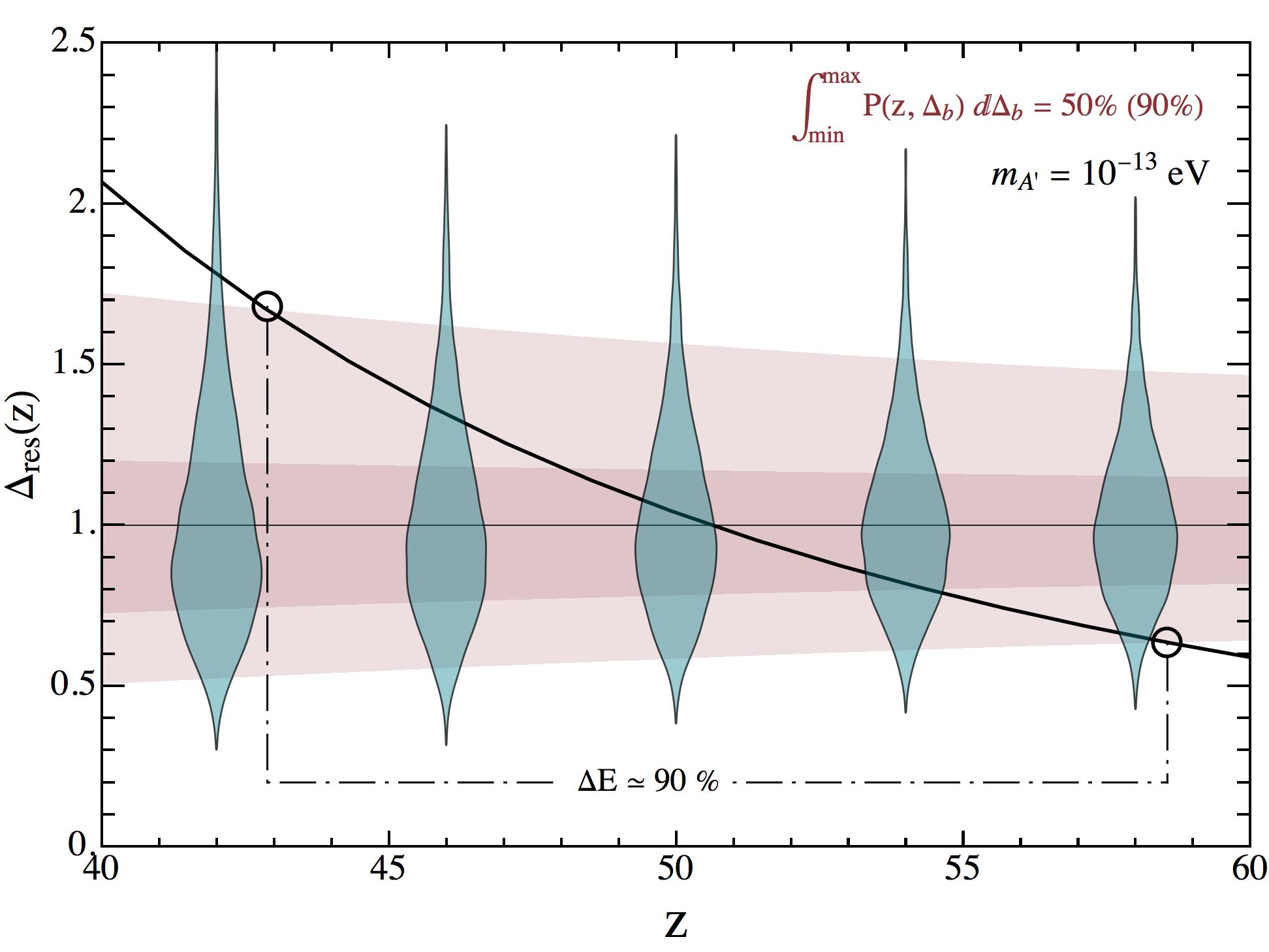}
	\caption{ \label{fig:fracE_violin} Evolution of $\Delta_{\rm res}(z)$ (black line) for a dark photon with $m_{A'} = 10^{-13}$ eV. We plot the log-normal overdensity PDF (blue regions) at $z = 42, 46, 50, 54,$ and 58, and we illustrate using red bands the intervals containing 50\% and 90\% of the PDF (assuming equal weight in each tail). Identifying the intersection of the $\Delta_{\rm res}$ line with the edges of a particular band, one can approximate the fraction of energy injected in a particular redshift interval. }
\end{figure*}

In this section we briefly illustrate the impact that inhomogeneities have on dark photon energy injection. Our primary goal is to provide the reader with an idea of the timescales over which energy injection takes place, as well as an idea of which dark photon masses are capable of undergoing resonance as structure begins to form.

In \Fig{fig:den_violin} we plot the resonant overdensity $\Delta_{\rm res}$ as a function of redshift for three dark photon masses. We assume that at redshifts $z > 12$ the free electron fraction of the Universe is consistent with the pre-reionization value ($\sim 2 \times 10^{-4}$) and at $z < 7$ the Universe is fully ionized (these boundaries are not so well-known, though current observations suggest that the Universe must be fully reionized by $z = 6$~\cite{Fan:2005es}, and likely could not have started before $z \gtrsim 12$~\cite{Heinrich:2016ojb,Hazra:2017gtx,Villanueva-Domingo:2017ahx}). We omit redshifts between $7 \lesssim z \lesssim 12$ due to the complicated nature of reionization (see \Sec{sec:reion} for further discussion). We overlay on figure \Fig{fig:den_violin} the log-normal PDFs for $\Delta_b$ evaluated at the redshift for which $\Delta_{\rm res} = 1$; that is, the relative width  of these distributions displays the relative value of the PDF (we stress that the absolute width of these PDFs in redshift has no physical meaning -- they simply represent 1-D PDFs valid at a single redshift). The broadening of the PDF at low redshifts is reflected by the vertical size of the PDFs. Using the scaling relation $\omega_p \propto \sqrt{\Delta_b}$, one can see that the PDF at small redshifts accommodates resonant transitions spanning roughly one order of magnitude in $m_{A'}$, while the one at $z\sim 30$ spans roughly a factor of two in $m_{A'}$. The lines in this diagram show the value of $\Delta_b$ required to achieve resonance with the dark photon of the mass indicated by that color. The location where these lines cross the dashed black line provides a visual solution to the redshift for which $\Delta_{\rm res} = 1$.   

In order to illustrate the rate at which energy is deposited for a particular model, we show in \Fig{fig:fracE_violin} the evolution of $\Delta_{\rm res}$ for $m_{A'}=10^{-13}$ eV. We overlay 5 PDFs $P_\Delta$ characterizing overdensities at $z = 42, 46, 50, 52,$ and 56. We highlight in red the regions of the PDFs which contain 50 and 90$\%$ of the weight, defined such that each of the tails contains the remaining $25$ and 5$\%$, respectively.  From these, one can estimate the timescale over which 50\% or 90\% of the energy is injected by finding the range over which the $\Delta_{\rm res}$ line overlaps a particular red region -- this is not exact as it neglects the redshift dependence of $\rho_{\rm CDM}(z)$ and $H(z)$,  and it does not account for the extra $\Delta_b$ weighting in \Eq{rhozdb}; this method serves only as a reasonable proxy for more exact solutions. This is illustrated using the $90\%$ interval in \Fig{fig:fracE_violin}, which shows that most of the energy is injected for redshifts $43 \lesssim z \lesssim 58$. This result is typical of models which have resonance in the range $10 \lesssim z \lesssim 50$. We have explicitly verified that dark photons undergoing resonance prior to reionization typically deposit $\sim 50\%$ of their energy in an interval $\pm \Delta z \sim 3$, while in the post reionization epoch this interval decreases to $\pm \Delta z \sim 1$, where in both cases a majority of the energy injection is centered around $z_{\rm res}(\Delta_b = 1)$. For dark photons that experience resonance at very late times, part of the energy injected ``should'' take place in the future, and thus the total energy injected is suppressed relative to the homogeneous case, but (for dark photon masses capable of resonantly converting in the homogeneous limit by today) this suppression is typically never much more than a factor of $2-3$. For masses larger than $10^{-13}$ eV the redshift range over which energy is deposited decreases, albeit quite slowly; this occurs because the dark matter and baryon power spectrum begin to diverge significantly by $z \sim 100$, with the baryon power spectrum having a significantly smaller variance than that of the dark matter. However, we do not attempt to quantify this effect precisely, as it does not have any significant impact for this study.

\section{Lyman-$\alpha$ Observations of the Epoch of HeII Reionization}\label{sec:lymanA}

Recent years have shown significant progress in the field of high-redshift Ly-$\alpha$ cosmology. Various analyses have shown that these observations, which indirectly probe the evolution of the temperature of the IGM during and after the epoch of reionization, are quite robust to astrophysical uncertainties (see \eg~\cite{McQuinn:2015icp} for a review). Exotic heating of the IGM during the post-reionization epoch can therefore be constrained to the level of $\leq 0.5$ eV / baryon  for $2 \lesssim z \lesssim 6 $~\cite{Sanderbeck:2015bba,onorbe2017self,Walther:2018pnn,Gaikwad:2020art,Bosman:2018xxh}.  Using this bound, Ref.~\cite{McDermott:2019lch} was able to constrain the kinetic mixing of dark photons undergoing resonance to be $\lesssim 2 \times 10^{-15}$; this bound, however, assumed a homogeneous Universe, which is not exact.  A more precise derivation of a bound based on Ly-$\alpha$ observations requires two primary modifications to account for the presence of inhomogeneities: one must account for the fact that ${\emph (i)}$ energy is injected inhomogeneously into the IGM, and ${\emph (ii)}$ Ly-$\alpha$ observations are not uniformly sensitive to all phases of the IGM~\cite{Becker:2010cu,Lukic:2014gqa}.  The former of these effects can be treated using the formalism derived in \Sec{sec:inhomoU}; we address the proper treatment of the latter effect below.

The Ly-$\alpha$ forest is an absorption phenomenon that occurs when light produced from distant QSOs (quasi-stellar objects, or quasars) passes through neutral hydrogen. Since the spectrum redshifts through the Ly-$\alpha$ frequency as it travels toward Earth, one can observe many Ly-$\alpha$ lines whose height and width characterize the properties of the IGM (\eg density, temperature, etc.). If the photons traverse large over-densities they will be preferentially absorbed, and in the extreme case of  $\sim 100\%$ absorption, no line will be observed here at Earth; consequently it is not possible to characterize the properties of over-densities responsible for near-total absorption. Conversely, photons traversing under-densities will easily pass through; if the transmission is $\sim 100\%$, one loses sensitivity to the properties of the IGM as well, since there is no line from which information can be extracted. Thus, one expects that the Ly-$\alpha$ spectrum observed at a given redshift is only sensitive to a finite range of inhomogeneities. To account for this, we assume that the sensitivity of the Ly-$\alpha$ observations to the temperature of the IGM is characterized by a function $\mathcal{S}(z,\Delta_b)$, and thus the energy injection which  Ly-$\alpha$ observations are sensitive to is given by
 \begin{equation}\label{eq:lymanE}
  \varepsilon_{\text{Ly-}\alpha} = \frac{\pi \epsilon^2 m_{A^\prime}}{3} \, \int \, d\Delta_b  \, \frac{{\rho}^{\rm homo}_{\rm CDM}\left(z_{\rm res} \right)}{{n}^{\rm homo}_b(z_{\rm res})H(z_{\rm res})}\, \mathcal{S}(z_{\rm res},\Delta_b)\,P_\Delta(z_{\rm res}, \Delta_b) \Theta(z_{\rm res} - 2) \Theta(6 - z_{\rm res}) \, .
  \end{equation}
It is the quantity in \Eq{eq:lymanE} that is constrained to be $\leq 0.5$ eV/baryon.

\begin{figure*}
	\includegraphics[width=0.49\textwidth]{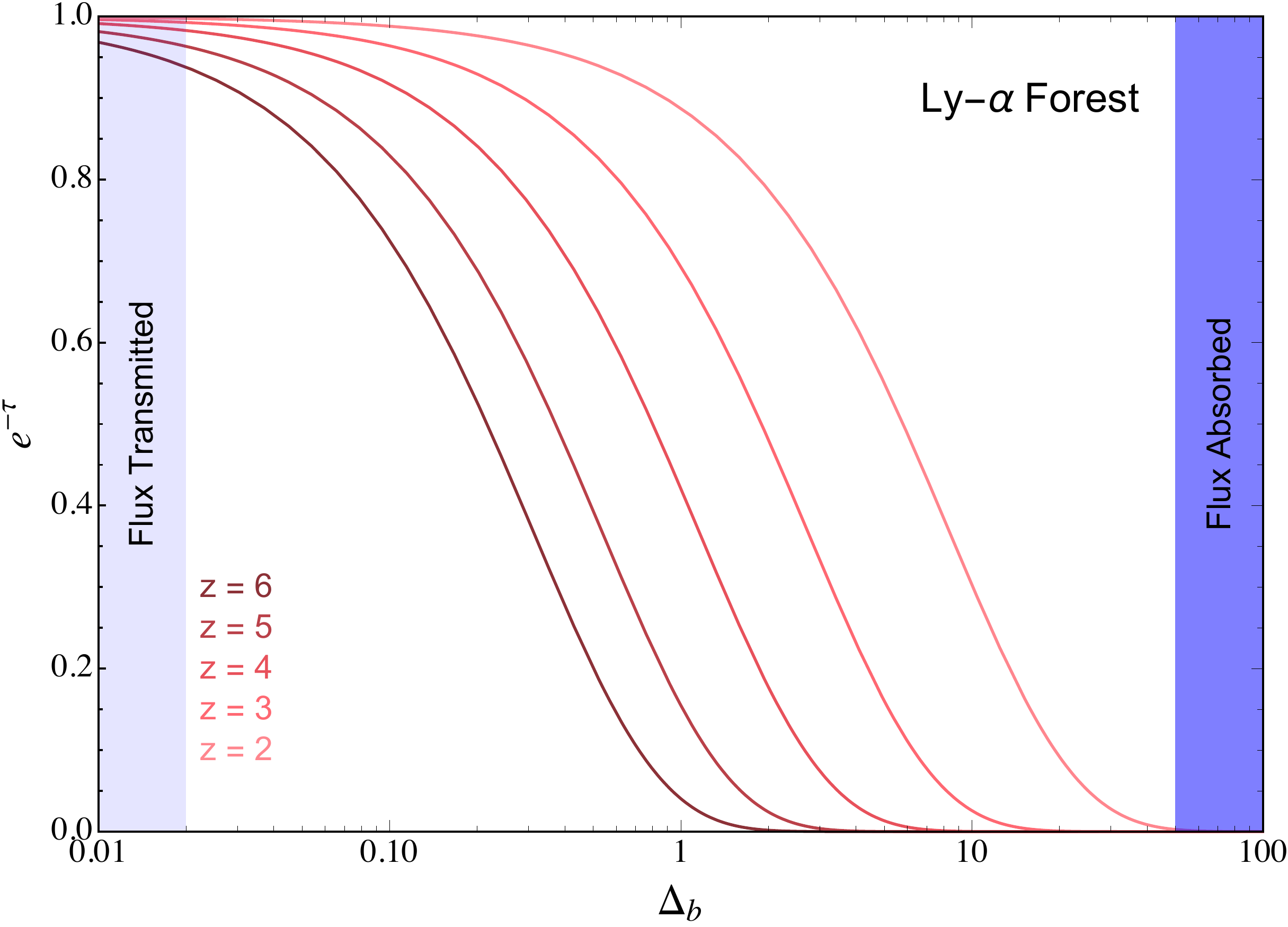}~
		\includegraphics[width=0.48\textwidth]{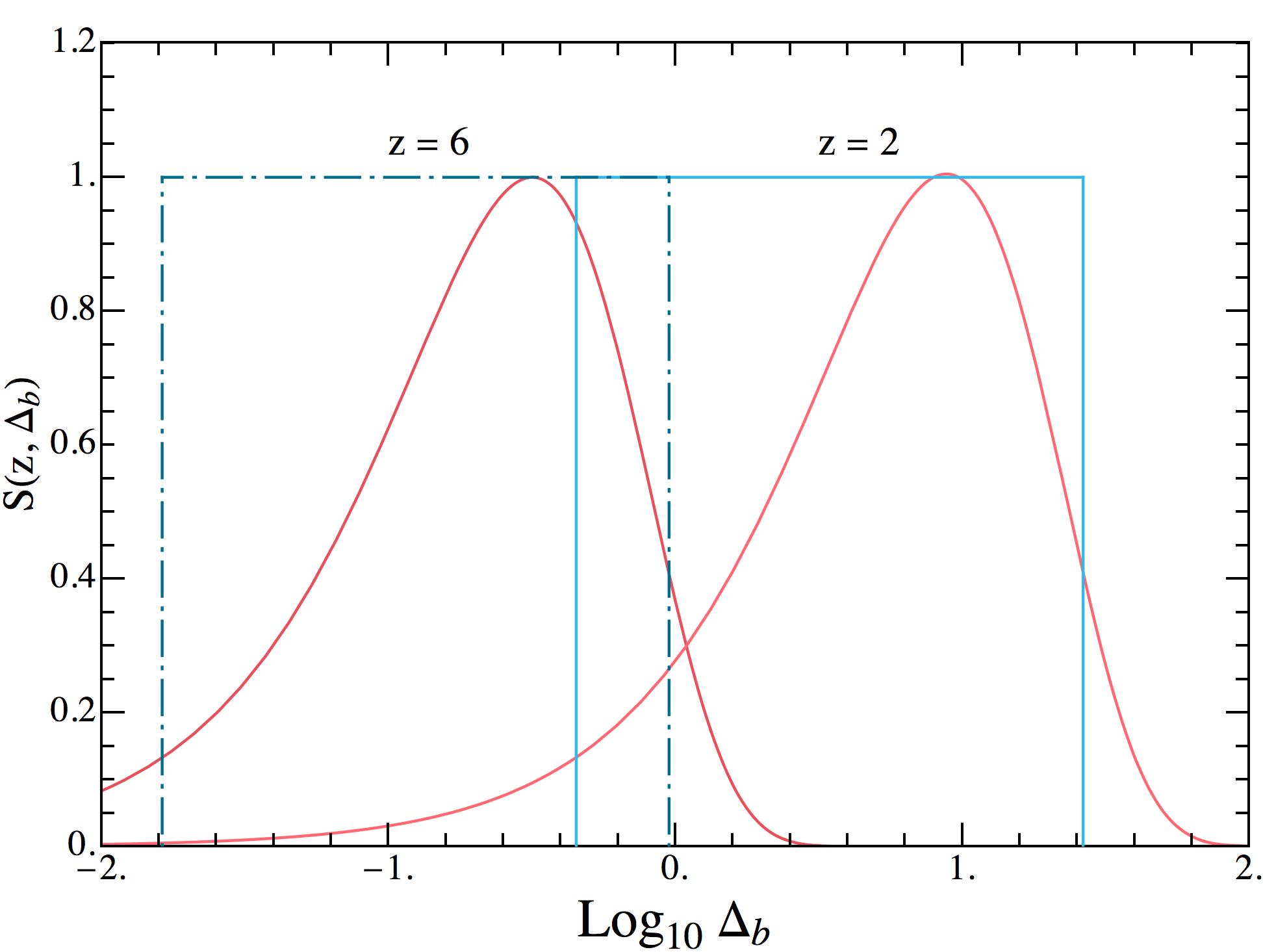}
	\caption{\label{fig:optD}Left: Absorption probability of Ly-$\alpha$ forest flux as computed in Ref.~\cite{Lukic:2014gqa}. Here, we have taken the redshift dependence of $T_0$ and $\gamma$, defined as $T(\Delta_b) \equiv T_0 \Delta_b^{\gamma - 1}$, from~\cite{Gaikwad:2020art} and used the measurements in Ref.~\cite{Becker:2010cu} to normalize the optical depth at each redshift.  Right: Adopted Ly-$\alpha$ sensitivity functions $\mathcal{S}(\Delta_b, z)$ from \Eq{eq:S1} (blue) and \Eq{eq:S2} (red) for $z = 2$ and $z = 6$. }
	\end{figure*}

We make two choices for the function $\mathcal{S}(z,\Delta_b)$ based on the absorption probability $e^{-\tau(z, \Delta_b)}$, where $\tau(z, \Delta_b)$ is the Ly-$\alpha$ optical depth.  First, we assume that the Ly-$\alpha$ spectrum is only sensitive to scales for which absorption is neither too large nor too small, \ie $\epsilon \leq e^{-\tau(z, \Delta_b)} \leq (1-\epsilon)$. Specifically, we adopt the following	
\begin{equation}\label{eq:S1}
	\mathcal{S}(z, \Delta_b) =  \begin{cases} 
 	0 & e^{-\tau(z, \Delta_b)} > 0.95 \\
 	1 & 0.05 \leq e^{-\tau(z, \Delta_b)} \leq 0.95 \\
 	0 & e^{-\tau(z, \Delta_b)} < 0.05 \, ,
 \end{cases}
\end{equation}
where the sensitivity thresholds are (roughly) based on the signal-to-noise ratio of Ly-$\alpha$ observations \cite{Gaikwad:2020art}. For an alternative approach, we adopt a sensitivity function that varies smoothly with overdensity. Since the extent to which the properties of the IGM can be extracted from Ly-$\alpha$ observations directly depends on the transmitted flux, it stands to reason that the sensitivity of a Ly-$\alpha$ observation on the thermal state of a particular over-density depends directly on the extent to which the absorption probability changes when the density field is varied (that is to say, Ly-$\alpha$ observations will be extremely sensitive to a particular inhomogeneity if small perturbations about that density $\Delta_b$ produce large changes in absorption probability). Concretely, this means adopting a sensitivity function proportional to the derivative of the absorption probability. In this case, we take
\begin{equation}\label{eq:S2}
	\mathcal{S}(z, \Delta_b) = A(z) \, \frac{\partial \tau(z, \Delta_b)}{\partial \Delta_b}  \, \Delta_b \, e^{-\tau(z, \Delta_b)} \, ,
\end{equation}
where the extra factor of $\Delta_b$ comes from the fact that we take the derivative with respect to the $\log \Delta_b$, and we choose the normalization $A(z)$ such that ${\rm Max}(\mathcal{S}(z)) = 1$. We illustrate the behavior of these sensitivity functions in the right panel of \Fig{fig:optD} for $z=2$ and $z=6$. Optimal over-densities for Ly-$\alpha$ observations in the interval $z\in[6,2]$ typically lie near $0.5 \lesssim \Delta_b \lesssim 6$~\cite{onorbe2017self}, which is in good agreement with the approximation adopted in \Eq{eq:S2}.

 In order to compute the sensitivity functions, we must estimate the Ly-$\alpha$ optical depth $\tau(z, \Delta_b)$. Generally speaking, the optical depth $\tau$ of photons to a given process is defined as
\begin{equation}
	\tau = \int d\ell \, n_x \sigma \, ,
\end{equation}
where $\sigma$ is the cross section for a photon to scatter from a species with number density $n_x$ along a path $d\ell$. The cross section for resonant line scattering is given by~\cite{Lukic:2014gqa}
\begin{equation}
	\sigma = \frac{\pi e^2}{m_e}f_{lu}\frac{1}{\Delta \nu_D} \phi_\nu \, ,
\end{equation}
where $f_{lu} = 0.416$ is the oscillator strength of the Ly-$\alpha$ transition, $\Delta \nu_D = b\,\nu_0$ is the doppler width with Doppler parameter $b = \sqrt{2 T_k / m_H}$  and central line frequency $\nu_0$, and $\phi_\nu$ is the line profile which we model here as the Gaussian core of a Voigt profile, \ie $\phi_\nu =  \exp(-x^2) / \sqrt{\pi}$, with $x = (\nu - \nu_0)/\Delta \nu_D$. Converting the line of sight integration to a redshift integration, we find that the optical depth can be expressed as
\begin{equation}
	\tau = \frac{\pi e^2 \, f_{lu} }{m_e b \, \nu_0} \, \int_0^{z_{\rm emit}} \, dz \, \frac{1}{H(z) \, (1+z)}\frac{\bar{n}_H}{\sqrt{\pi} \, \Delta \nu_D} \, e^{- \left(\frac{\nu - \nu_0}{\Delta \nu_D} \right)^2} \, ,
\end{equation}
where $\nu$ is photon frequency at a particular redshift, and $\bar{n}_H$ is the neutral hydrogen along the line of sight (which is proportional to $\Delta_b$). Since the width of the line profile is extremely narrow, the absorption is dominated by a narrow redshift region near the source. For example, given that typical Doppler parameters are on the order $b \sim 10^{-4}$, the exponential term for a photon which has the Ly-$\alpha$ frequency at $z = 6$ will have dropped by a factor of $\sim 3$ by a redshift $z = 5.9993$. Consequently, we can approximate the entire contribution as being local, \ie
\begin{equation}
	\tau \simeq \frac{\pi e^2 \, f_{lu} }{m_e \, \nu_0} \,  \frac{ \bar{n}_H}{2 H(z_{})} \, {\rm Erf}\bL \frac{z_{}}{b (1+z)}\bR \, .
\end{equation}
 Following~\cite{Sanderbeck:2015bba,Gaikwad:2020art,sanderbeck2020inhomogeneous}, we assume that the temperature of an over-density scales like $T = T_0 \Delta_b^{\gamma - 1}$, where $T_0$ is the temperature at the mean density and $\gamma \sim 1.5$~\cite{Sanderbeck:2015bba,Gaikwad:2020art,sanderbeck2020inhomogeneous}.  This problem is complicated by the fact that the average local neutral hydrogen fraction $\bar{n}_H$ is not a known quantity; to leading order the Universe is fully ionized at these redshifts, and thus na\"ively $\bar{n}_H \sim 0$. This issue can be resolved using the method described in Refs.~\cite{Becker:2010cu,Lukic:2014gqa},
 which relies on renormalizing $\tau(z)$ to ensure observations match the results of hydrodynamical simulations. In other words, this unknown normalization can be  determined by defining an effective optical depth $\tau_{\rm eff}$ via
\begin{equation}
	\tau_{\rm eff}(z) = \int d\Delta_b \, P_\Delta(z,\Delta_b) \, \tau(z,\Delta_b) \, ,
\end{equation}  
and using the result of Ref.~\cite{Becker:2010cu} to fix \eg $\tau_{\rm eff}(z=1.9) \sim 0.1$.

We illustrate the behavior of the suppression factor $e^{-\tau}$ in \Fig{fig:optD}, where we plot the absorption probability of Ly-$\alpha$ photons as a function of overdensity for various redshifts. Interestingly, \Fig{fig:optD} shows that $\exp(-\tau) \lesssim 0.2$ for redshifts $z \gtrsim 5$  when $\Delta_b \gtrsim 1$, and thus that Ly-$\alpha$ observations are effectively insensitive to energy injection in halos. We conclude that constraints on dark photons with masses larger than $m_{A'} \simeq 4\tenx{-13}\ev$ or much smaller than $5 \times 10^{-14}\ev$ will be heavily suppressed.

\begin{figure*}
	\includegraphics[width=0.49\textwidth]{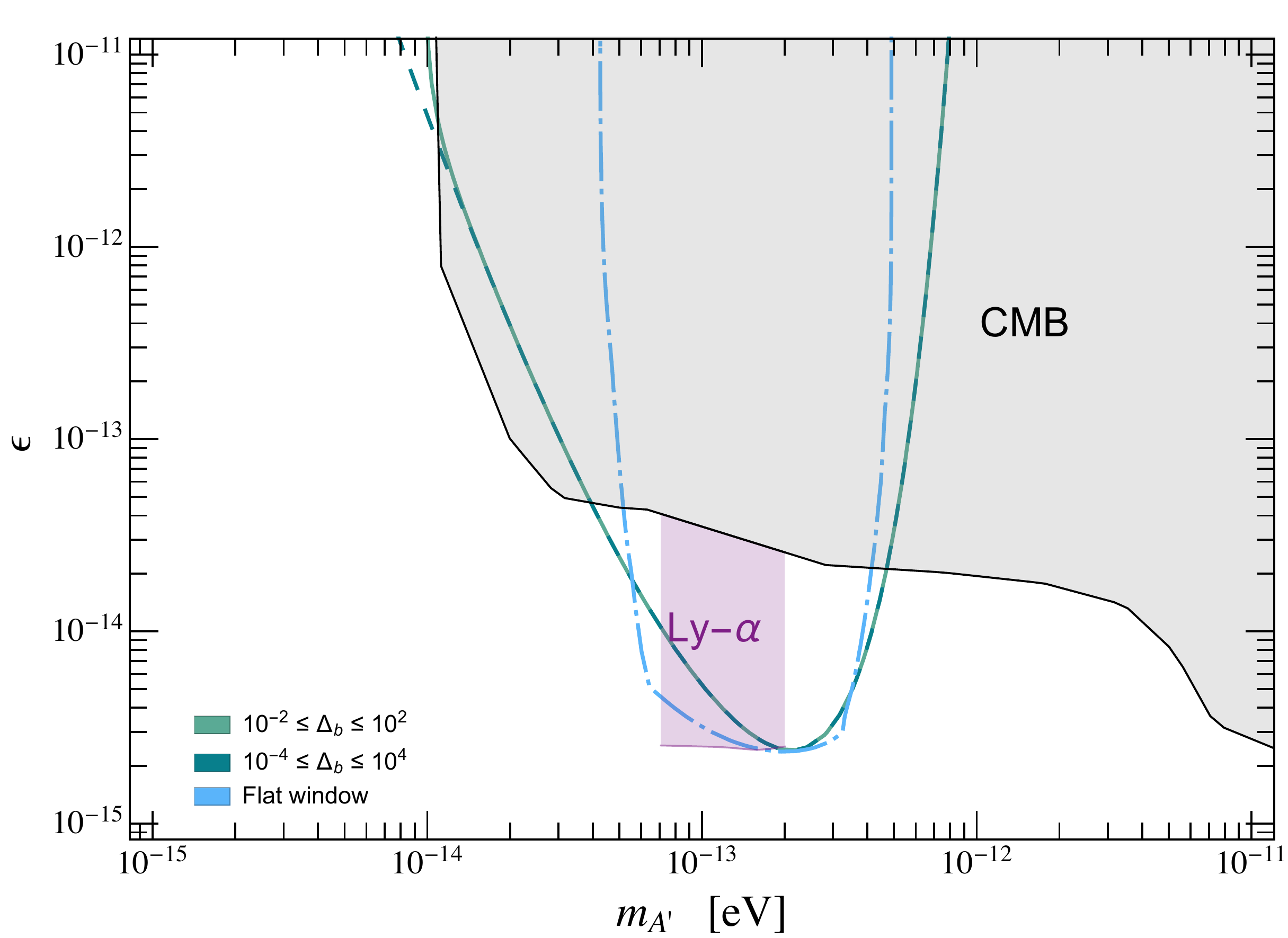}
	\includegraphics[width=0.49\textwidth]{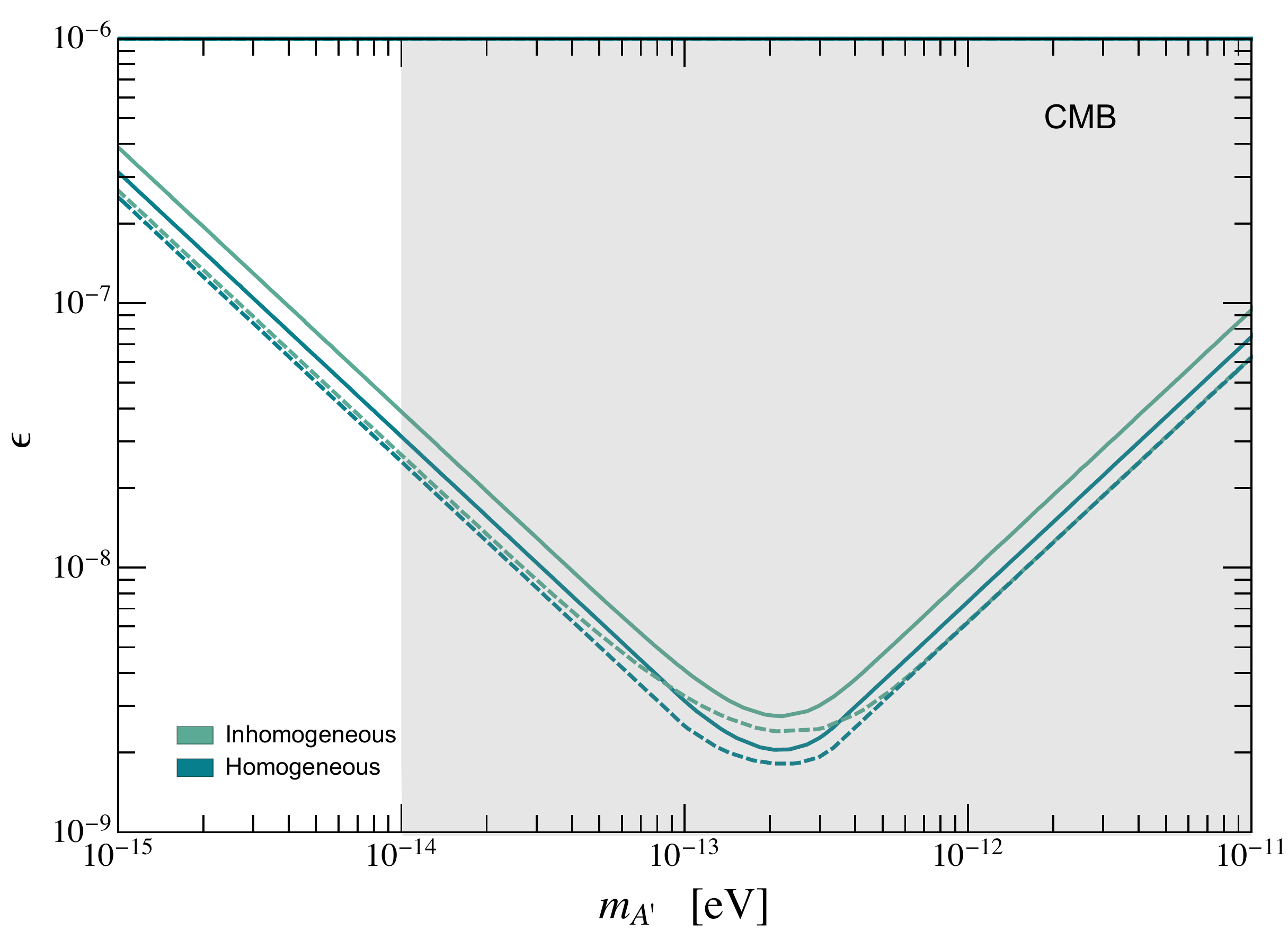}
	\caption{\label{fig:lyBND} Left: Constraints from Ly-$\alpha$ observations between $2 \lesssim z \lesssim 6$; obtained by requiring $\varepsilon_{\text{Ly-}\alpha}<0.5$ eV/baryon, as defined in \Eq{eq:lymanE}. Constraints are derived assuming the PDF of $\Delta_b$ is log-normal and is valid in the range $10^{-2} \leq \Delta_b \leq 10^2$ or $10^{-4} \leq \Delta_b \leq 10^4$, and using two different sensitivity functions as divided in \Eq{eq:S1} (blue) and \Eq{eq:S2} (green). These constraints are compared with those derived from the CMB (grey) and Ly-$\alpha$ forest (purple) assuming a homogeneous Universe (as derived in Ref.~\cite{McDermott:2019lch}).  Right: Non-resonant constraints derived with and without inhomogeneities, and including (solid) and excluding (dashed) the default sensitivity function in \Eq{eq:S2}. Grey region denotes parameter space excluded using the CMB~\cite{McDermott:2019lch}.}
	\end{figure*}

Using \Eq{eq:lymanE} and \Eq{eq:inhomIB}, we extend the constraints obtained in Ref.~\cite{McDermott:2019lch} to account for the presence of inhomogeneous structure, both for resonant and non-resonant absorption. This is done by requiring that the observable energy injected given in \Eq{eq:lymanE} is equal to 0.5 eV / baryon in the interval $z \in [2, 6]$. These constraints are shown in \Fig{fig:lyBND}\footnote{We do not show bounds derived from black hole superradiance~\cite{Baryakhtar:2017ngi}, which in principle can be used to constrain ultralight dark photons with masses near this range, as the existence of such bounds is model dependent~\cite{Agrawal:2018vin}.}, assuming the sensitivity functions are given as in \Eq{eq:S2} (assumed to be the default sensitivity) and \Eq{eq:S1} (dubbed `Flat window'). For the default sensitivity function, we also show the effect of increasing the integration from $10^{-2} \leq \Delta_b \leq 10^2$ to $10^{-4} \leq \Delta_b \leq 10^4$; the difference is negligible everywhere except the low-$m_{A'}$ tail. Note that Ref.~\cite{Caputo:2020bdy} recently performed a similar analysis, however reached a rather different result -- we make a detailed comparison of the two approaches in \App{sec:comparison}. In the right panel of \Fig{fig:lyBND} we plot the bounds from non-resonant absorption in both the homogeneous and inhomogeneous limit, assuming either $\mathcal{S}(z, \Delta_b) = 1$ (dashed) or is given by \Eq{eq:S2}. We see that the effect of structure and the sensitivity function have a minimal effect on the net sensitivity of the non-resonant absorption bounds. 

\section{21cm Cosmology}\label{sec:21cm}

The field of 21cm cosmology aims to understand the spatiotemporal evolution of neutral hydrogen in the Universe by studying the hyperfine transition between the ground and first excited state (this is also known as the ``21cm'' transition). The goal, simply put, is to measure the evolution of the intensity of the redshifted 21cm line; the frequency of an observed line provides direct information on the epoch at which the absorption or emission took place, and the intensity provides various pieces of information on the state of neutral hydrogen, such as its temperature, number density,  line-of-sight velocity, and ionization fraction. We review the basics of 21cm cosmology below, but refer the interested reader to Refs.~\cite{Madau:1996cs,Furlanetto:2006jb,Pritchard:2011xb} for more extensive reviews.

The amount of absorption or emission of the gas is determined by the relative occupation number of the ground and excited states, a quantity which is typically parameterized with the so-called spin temperature $T_s$, defined via
\begin{equation}
	\frac{n_1}{n_0} = 3 \, e^{- T_{21} / T_s} \, ,
\end{equation}
where $n_1$ and $n_0$ represent the excited and ground state, the factor of 3 is the degeneracy factor the excited state, and $T_{21} \equiv h \nu_{21} / k_B$. Focusing our attention on neutral hydrogen in the IGM during the dark ages, only a small number of processes are capable of changing the ratio of ground and excited states, or equivalently of changing the spin temperature. These are: spontaneous emission, stimulated absorption/emission, collisional excitation/de-excitation, and indirect excitation/de-excitation via Ly-$\alpha$ pumping. The spin temperature can be expressed in terms of effective coupling coefficients $y_i$ that characterize the efficiency of each of these processes, the temperature of the background radiation $T_r$, and the (kinetic) temperature of matter $T_k$, as
\begin{equation}\label{eq:ts}
	T_s \simeq \frac{T_R + (y_k + y_\alpha) T_k }{1 + y_k + y_\alpha} \, .
\end{equation}
The collisional and Ly-$\alpha$ couplings ($y_k$ and $y_\alpha$, respectively) are positive semi-definite quantities whose definitions can be found in~\eg Ref.~\cite{Mena:2019nhm}. 

Radio experiments searching for the redshifted 21cm line are only sensitive to the relative difference between the intensity produced by these transitions and radio background,  which is expected to be dominated by photons in the Rayleigh-Jeans tail of CMB blackbody. As such, the signal is typically expressed as a differential measurement relative to the CMB intensity. Following the convention in radio astronomy to work with effective brightness temperatures $T_b$ rather than intensity itself (the two are directly related by the Rayleigh-Jeans limit of the blackbody relation), the 21cm signal (\ie the differential brightness temperature) can be expressed as
\begin{equation}
	\delta T_b (\nu) = \frac{T_s - T_{\rm cmb}}{1+z}(1 - e^{-\tau_{\nu_0}}) \, ,
\end{equation}
where $\tau_{\nu_0}$ is the optical depth of the 21cm line. Since the optical depth is small, the exponential term can be expanded, yielding a differential brightness temperature along a line of sight equal to
\begin{equation}\label{eq:gentb}
	\delta T_b \simeq 27 x_{HI} \sqrt{\frac{z+1}{10}} \Delta_b \left(1 - \frac{T_{\rm cmb}}{T_s} \right) \, {\rm mK} \, , 
\end{equation}
where $x_{HI}$ is the neutral hydrogen fraction. It is clear in \Eq{eq:gentb} that the differential brightness temperature will be seen in emission (\ie $\delta T_b > 0$) if $T_s > T_{\rm  cmb}$ and absorption (\ie $\delta T_b < 0$) if $T_s < T_{\rm cmb}$.  Furthermore, from \Eq{eq:ts} one can  see that the spin temperature is bounded to be between the temperature of the CMB and the matter temperature. An immediate consequence is that, if the 21cm signal is observed in absorption (as is expected for $z\gtrsim 15$), one can immediately infer that $T_k < T_{\rm cmb}$  and constrain the maximum temperature of the IGM by taking $T_s \to T_k$ in \Eq{eq:gentb}.  We devote the remainder of this section to investigating the extent to which future 21cm experiments can constrain the heating induced by dark photons should they observe the 21cm signal in absorption.

The simplest and least expensive  21cm experiments are comprised of single antennas that attempt to measure the sky-averaged differential brightness temperature. Recently, the EDGES collaboration~\cite{Bowman:2018yin} claimed the first detection of the global 21cm differential brightness temperature. At the moment, the validity of this measurement is still a matter of hot debate due the complicated nature of foreground removal and its incompatibility with the $\Lambda$CDM prediction. Consequently, in this work we will assume that the true 21cm differential brightness temperature is as yet unknown. In the near future, radio interferometers such as HERA\footnote{\url{https://reionization.org/}}~\cite{Beardsley:2014bea,DeBoer:2016tnn} and SKA\footnote{\url{https://www.skatelescope.org/}}~\cite{Mellema:2012ht} will measure the 21cm power spectrum from reionization to redshifts as high as $z\lesssim 25$. Radio interferometers have a clear advantage over single dish antennas in that smooth radio backgrounds contribute only to low multipoles, which are easy to remove. In addition, the power spectrum contains far more information than the differential brightness temperature, and thus can be a more powerful tool in constraining exotic physics. Incorporating the effect of inhomogeneous energy injection from dark photons in this case is, however, rather involved, and thus we postpone this to future work.  In this study, we instead focus on the globally averaged differential brightness temperature $\left< \delta T_b \right>$ (similar sensitivity estimates have been made using the average differential brightness temperature in the homogeneous limit in Ref.~\cite{Kovetz:2018zes}).

Generalizing \Eq{eq:gentb}, we can account for the presence of inhomogeneities in the globally averaged differential brightness temperature~\cite{Villanueva-Domingo:2019ysf}:
\begin{equation}
	\left< \delta T_b\right> = 27 x_{HI} \sqrt{\frac{z+1}{10}} \bL 1 - \pL \int d\Delta_b \, P_\Delta(z,\Delta_b) \,\Delta_b\frac{T_{\rm cmb}(z)}{T_s(\Delta_b, z)} \pR \bR \, {\rm mK} \, ,
\end{equation}
where we have included the explicit dependence of the spin temperature on the redshift and over-density. Since we are interested in determining the maximal level of absorption, we can make the substitution $T_s(\Delta_b, z) \to T_k(\Delta_b, z)$, which in most cases will be an overly conservative estimate (see \eg Ref.~\cite{Witte:2018itc} to understand the expected contribution of x-ray heating by stellar sources). We can then solve for the evolution of the matter temperature at each possible overdensity using 
\begin{equation}\label{eq:tk_evol}
	\frac{dT_k}{dt} + 2 H T_k -  \frac{2}{3}\frac{T_k}{\Delta_b}\frac{d \Delta_b}{dt} + \frac{T_k}{1+x_e}\frac{x_e}{dt} = \frac{2 Q_{\rm inj}}{3 n_b (1+x_e + f_{\rm He})}\, ,
\end{equation}
where $Q_{\rm inj}$ is the heating rate per unit volume (which includes \eg x-ray heating and Compton cooling as well as dark photon heating).  We adopt initial conditions $T_k(\Delta_b) = \overline{T_{\rm ad}}\Delta_b^{2/3}$ (which reproduces the solution that when only adiabatic cooling is relevant~\cite{Villanueva-Domingo:2019ysf}), where $\overline{T_{\rm ad}}$ is the mean adiabatic temperature. In solving \Eq{eq:tk_evol} we neglect the term proportional to $d x_e/dt$, since we are focusing on the epoch prior to reionization where the free electron fraction is slowly changing. We also neglect the term proportional to $d \Delta_b / dt$, as this term for typical overdensities is expected to be small relative to to the contribution of adiabatic cooling. We do explicitly include the Compton cooling contribution in the right hand side of \Eq{eq:tk_evol}, as well as the exotic energy injection from resonant conversion. 

\begin{figure}
\includegraphics[width=0.495\textwidth]{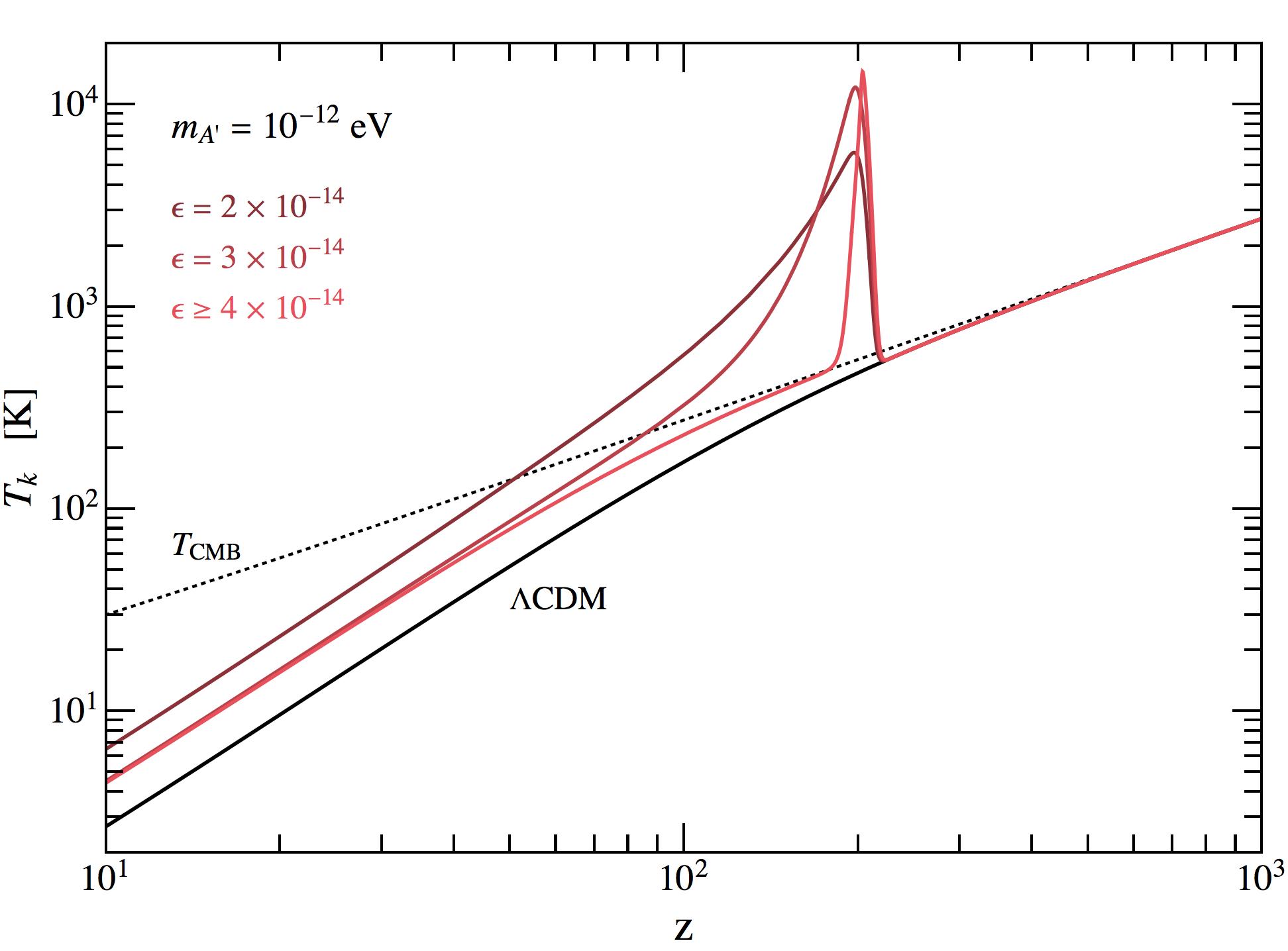}
\includegraphics[width=0.495\textwidth]{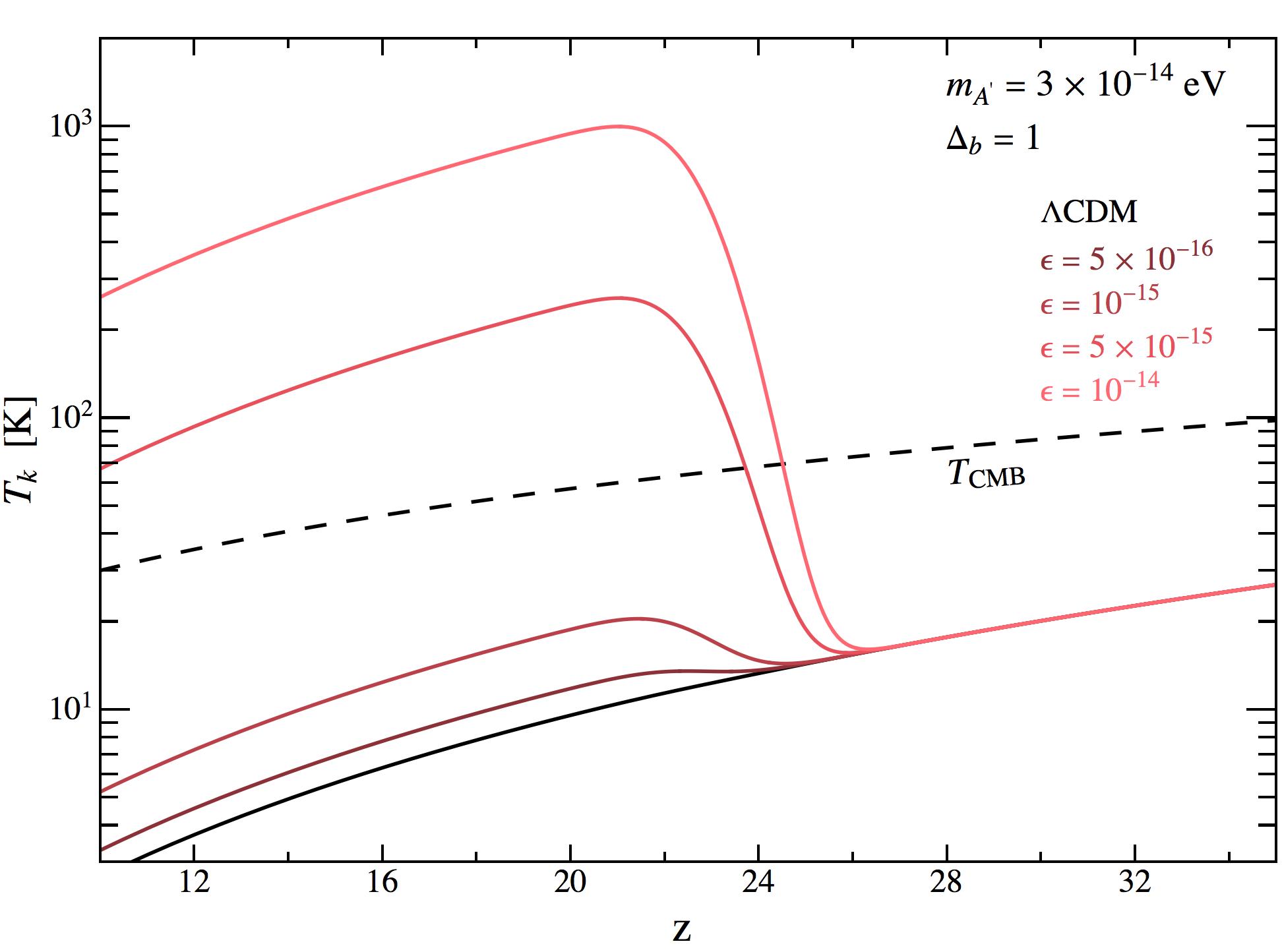}
\caption{Left: Evolution of matter temperature $T_k$ assuming homogeneous energy injection at high $z$ (modeled using Gaussian with a width $\Delta z= 5$). Right: Evolution of matter temperature at low redshift, assuming $\Delta_b = 1$, for $m_{A'}=3\tenx{-14}\ev$ and various values of $\ep$.}
\label{fig:matterT}
\end{figure}

In \Fig{fig:matterT} we show the evolution of matter temperature relative to that of $\Lambda$CDM and to the CMB temperature for dark photons which undergo resonance at high (left) and low (right) redshift. For the high redshift resonance, we solve this using {\tt Recfast++}~\cite{Seager:1999bc,Chluba:2010ca} (including a contribution from collisional ionization as in Refs.~\cite{matsuda1971dissipation,dopita2013astrophysics,McDermott:2019lch}), while at low redshift we solve \Eq{eq:tk_evol} for $\Delta_b = 1$. For high redshift resonances (left panel), it is difficult to significantly elevate the matter temperature above that of the CMB during the epoch for which 21cm observations will soon exist. This is because if one heats the medium above the threshold for collisional ionization $T_k\sim10^4$K, the cooling rate of the medium changes and the net heating saturates. The right panel shows that at late times the story is quite different. One can easily heat the medium above the CMB without encountering any issue with the ionization threshold.

\begin{figure*}
	\includegraphics[width=0.6\textwidth]{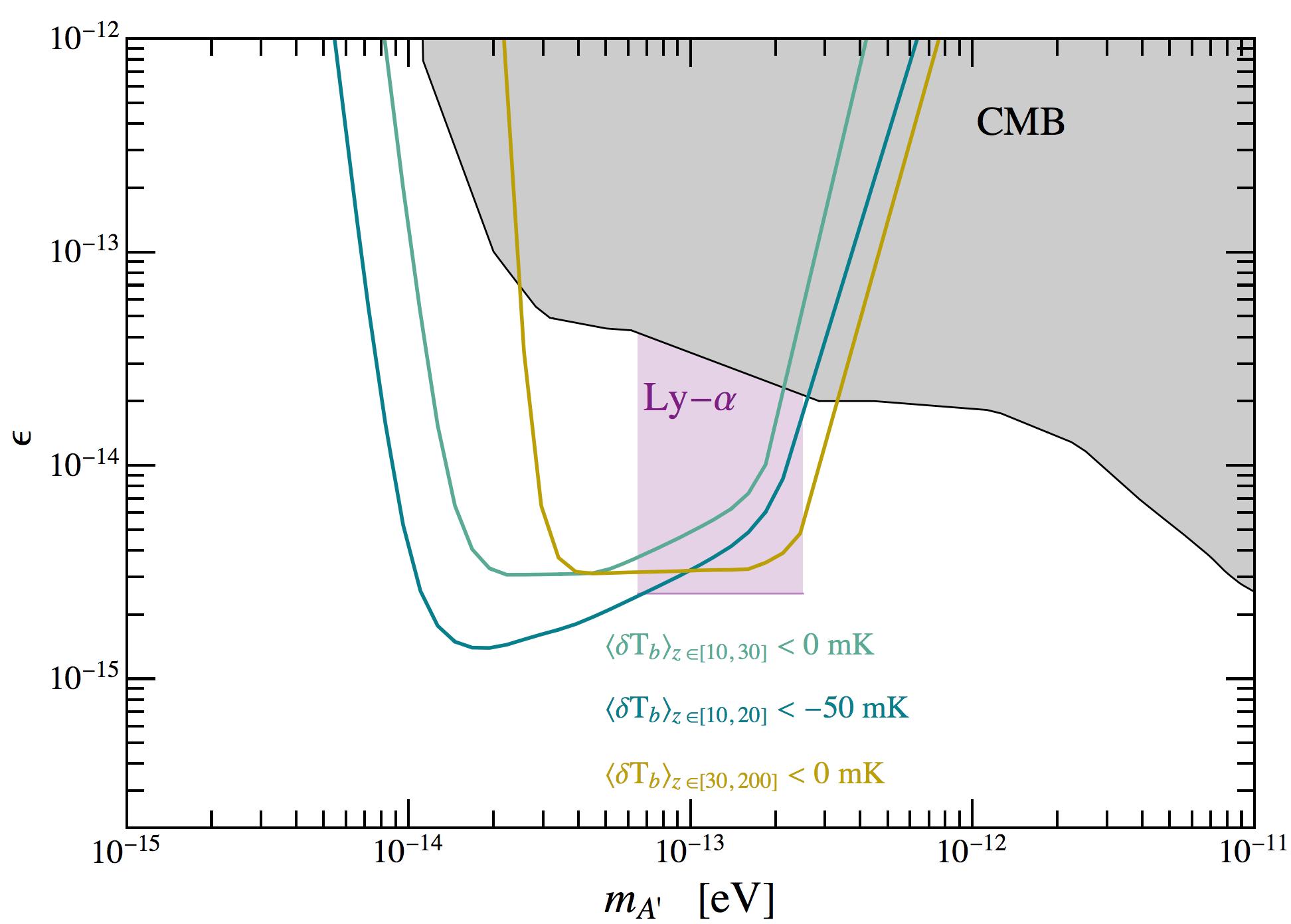}
	\caption{\label{fig:21cmBnd} Parameter space that could be excluded should an experiment observe the sky-averaged 21cm temperature in an interval $z \in [z_{\rm min}, z_{\rm max}]$ with an amplitude less than or equal to some threshold (being here $0$ mK or $-50$ mK).  }
\end{figure*}

In order to project potential sensitivity of 21cm experiments measuring the global brightness temperature to dark photon dark matter, we adopt two potential experimental configurations, one consistent with an SKA-like experiment~\cite{Mellema:2012ht} and the other being a lunar-based radio array~\cite{Burns:2011wf,Greenhill:2012mn,Koopmans:2019wbn}. We assume that these experiments measure absorption at the level of $\left< \delta T_b \right> \leq -50$ mK or $\leq 0$ mK across a range of redshifts. The model is assumed to be falsifiable if the dark photon heats the medium above this level at all points in this redshift range. In reality, constraints may be significantly stronger than this, but such a statement would rely on complicated astrophysical modeling at high redshift where little is currently known. Potential future
sensitivity are illustrated in \Fig{fig:21cmBnd}. This shows that 21cm observations could be extremely useful in extending sensitivity to lower-mass dark photons. 

Finally, it is worth mentioning that near-future radio interferometers will hope to move well beyond the globally averaged differential brightness temperature and measure the 21cm power spectrum.  Far more information is contained in the evolution of the power spectrum than in the global signal, particularly when energy injection proceeds in a largely inhomogeneous way. Consequently, radio interferometers will provide a great opportunity to probe the existence of dark photon dark matter.

\section{Discussion}\label{sec:discussion}
In this section we comment on a number of additional signatures and features that may appear in this model. In particular we discuss ${\emph (i)}$ the expectations for resonant conversion during the epoch of reionization,   ${\emph (ii)}$ potential implications for star formation rates,  ${\emph (iii)}$ expected modifications due to bounds derived from the CMB optical depth, and  ${\emph (iv) }$ late-time $y$-type spectral distortions due to excess heating of the IGM.

\subsection{Reionization}\label{sec:reion}

The epoch of reionization refers to the period during which the Universe evolved from being predominately neutral to fully ionized. Measurements of the Gunn-Peterson trough in the spectrum of quasars provide convincing evidence that reionization had completed by $z \sim 6$~\cite{Fan:2005es}. On the other hand, the integrated optical depth of the CMB measured by Planck \cite{Aghanim:2018eyx} does not allow for significant changes to the free electron fraction at $z \gtrsim 15$~\cite{Heinrich:2016ojb,Hazra:2017gtx,Villanueva-Domingo:2017ahx}. Between these epochs, UV radiation (likely) sourced from the first collapsed objects is expected to permeate the Universe and rapidly change the free electron fraction. Understanding how this process began and evolved is currently an active area of research. Still, there are some features which appear consistent among leading theories: namely, the ionizing photons were likely produced in over-densities and had relatively short mean free paths in the neutral media that they were ionizing. Consequently, to first order one can imagine that reionization proceeded via the formation of ionized bubbles, inside of which the Universe is fully ionized (with $x_e \sim 1.08$), while the IGM outside the bubbles remained nearly unaffected. The free electron fraction outside of the ionized bubbles would be given by the pre-reionization value $x_e \sim 10^{-4}$~\cite{Furlanetto:2005xx,Zahn:2006sg,McQuinn:2006et,Mesinger:2007pd}, until the bubbles grow and merge, at which point reionization is complete.

The dramatically inhomogeneous nature of reionization has implications for resonant dark photon conversion, since the inhomogeneous structure of $x_e(\vec{x}, z)$ complicates the spatial and temporal understanding of the resonance. Naively, one may be tempted to use the globally averaged value of the free electron fraction to understand which dark photon masses allow for resonant conversion during the reionization epoch. Unfortunately this is wrong since reionization is an intrinsically inhomogeneous process. A better attempt at treating the Universe during this epoch is to work in a two-phase approximation, \ie  assuming the co-existence of a fully ionized medium with a medium whose ionization level is consistent with the standard background evolution (without reionization sources), and neglect the boundaries between ionized and non-ionized regions. We believe that this is a reasonable approximation given the short mean free path of the ionizing photons in the neutral medium. The question then becomes: how should one map over-densities to ionized and non-ionized regions?

Here, we briefly sketch two possibilities for approximating the effect of energy injection from resonant dark photon conversion during this epoch. However, given the large theoretical uncertainties associated to modeling reionization, the constraints obtained in this section should not be interpreted as robust. Instead, this discussion is intended merely as an exercise to illustrate how considering inhomogeneities in $x_e$ during the epoch of reionization might affect the bounds derived elsewhere.

\begin{figure}
	\includegraphics[width=0.6\textwidth]{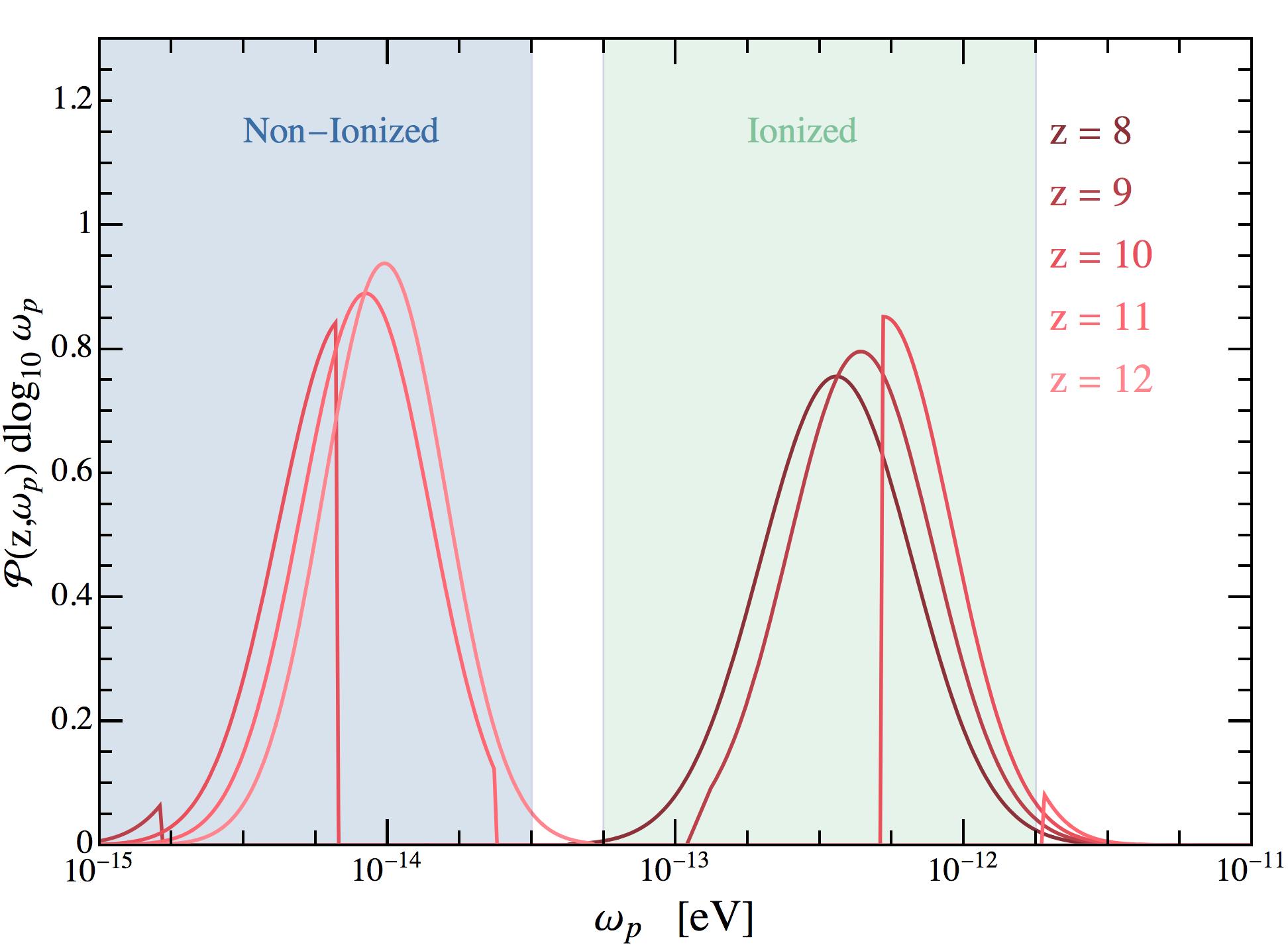}
	\caption{\label{fig:reion} Probability distribution function characterizing distribution of plasma frequencies in the IGM during reionization, assuming the `strongly inhomogeneous' scenario (in which reionization is assumed to progress from over-densities to under-densities). The left region captures the evolution of the predominantly neutral medium $x_e \sim 10^{-4}$, while the right region shows the evolution of the predominantly ionized regions $x_e \sim 1.08$. The mean free electron fraction is assumed to be a $\tanh$ function centered at $z =10$ with width $\Delta z = 0.5$. }
\end{figure}

Adopting the two phase approximation, and assuming we have a measurement of the global free electron fraction, one can estimate what volumetric fraction $V_{\rm ion}$ of the Universe is ionized  via 
\begin{equation}
	\overline{x_e} \sim (1+f_{\rm He}) V_{\rm ion}  + x_{e, {\rm pre-ion}} (1 - V_{\rm ion}) \, ,
\end{equation}
where $f_{\rm He} \sim 0.08$ and $x_{e, {\rm pre-ion}}$ is the value of the free electron fraction assuming ionization has not modified the evolution of the IGM.  The simplest assumption one can make is that under-densities and over-densities are equally likely to be ionized.
This amounts to a differential rate of energy injection of
\begin{equation}
	\frac{d \rho_{A^\prime \to \gamma}}{dz}(z) = V_{\rm ion} \,\frac{d \rho_{A^\prime \to \gamma}}{dz}(z)\Big|_{x_e = 1.08} + (1 - V_{\rm ion} )\, \frac{d \rho_{A^\prime \to \gamma}}{dz}(z) \Big|_{x_e = x_{e, {\rm pre-ion}}} \, .
\end{equation}
Here, the notation $|_{x_e = \cdots}$ means that the resonant redshift $z_{\rm res}$ in \Eq{eq:ZRES} is determined with a particular value of the free electron fraction.
We will refer to this scenario as the `homogeneous' reionization. As previously mentioned, however, we expect ionizing photons to be produced in collapsed objects, meaning they originate from over-densities and expand outwards. At the converse extreme, we might thus expect ionization to first occur in the densest objects, and proceed from over-densities to under-densities. We will refer to this scenario as the `strongly inhomogeneous' scenario. In this case, we can identify at each redshift the threshold of over-densities $\Delta_{\rm thresh}$ defining the boundary between ionized and non-ionized regions by solving
\begin{equation}
	V_{\rm ion} \simeq \int_{\Delta_{\rm thresh}} \, d\Delta_b\, P_\Delta(z, \Delta_b) \hspace{.4cm} {\rm or} \hspace{.4cm} (1 - V_{\rm ion}) \simeq \int^{\Delta_{\rm thresh}} \, d\Delta_b \,P_\Delta(z, \Delta_b) \, .
\end{equation}
In this case, one can write the differential rate of energy injection 
\begin{eqnarray*}
	\frac{d \rho_{A^\prime \to \gamma}}{dz}(z) &= {\rho}^{\rm homo}_{\rm CDM}(z)  \, \left[ \int^{\Delta_{\rm thresh}} d\Delta_b \, P_\Delta(z,\Delta_b) \, \Delta_b \, \frac{d}{dz} P_{A^\prime \rightarrow \gamma}(z, \Delta_b)\Big|_{x_e = x_{e, {\rm pre-ion}}}   \right. \\ & \left. +  \int_{\Delta_{\rm thresh}} d\Delta_b \, P_\Delta(z,\Delta_b) \, \Delta_b \, \frac{d}{dz} P_{A^\prime \rightarrow \gamma}(z, \Delta_b) \Big|_{x_e = 1.08}\right] \, .
\end{eqnarray*}
An important consequence of the strongly inhomogeneous scenario is the appearance of a gap in the evolution of the PDF characterizing the plasma frequency of the Universe. This is illustrated in \Fig{fig:reion}, where we plot the evolution of the PDF characterizing $\omega_p$ during reionization in the strongly inhomogeneous scenario, assuming the mean free electron fraction is given by a  $\tanh$ function centered at $z = 10$ and of width $\Delta z = 0.5$. We have highlighted the two phases via the labels `non-ionized' and `ionized'. The sharp cut in the distributions illustrates the evolution of $\Delta_{\rm thresh}$ with time.  

\begin{figure}
	\includegraphics[width=0.6\textwidth]{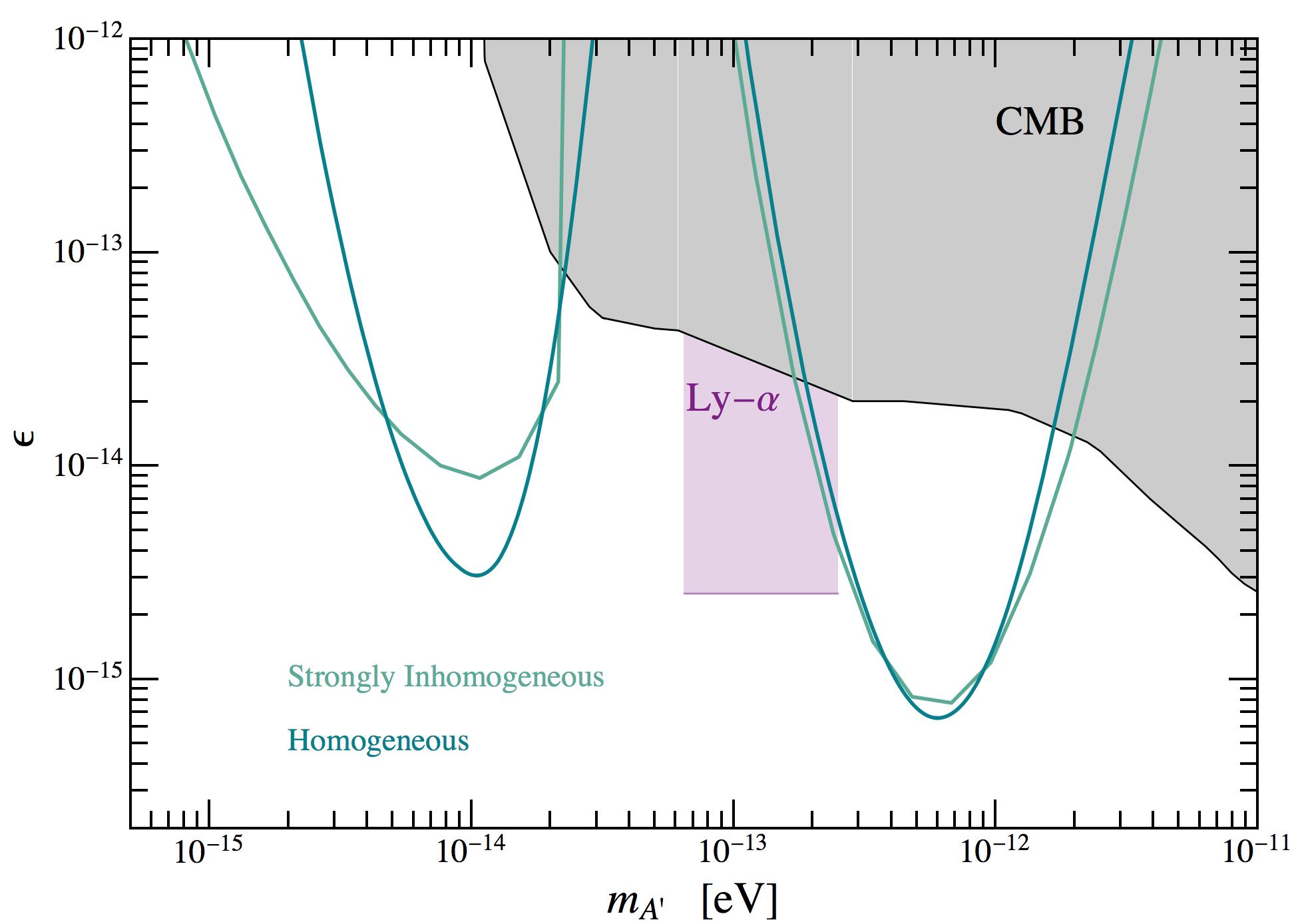}
	\caption{\label{fig:Rebnd} Bound that could be derived assuming $\varepsilon_{\rm inj}$ between $z \in [6,10]$ is less than $1$ eV / baryon, and assuming reionization proceeded either homogeneously (\ie all $\Delta_b$ ionized at same rate), or in a strongly inhomogeneous manner (ionizing over-densities first, and under-densities last). }
\end{figure}

For both reionization scenarios discussed, we derive an illustrative `bound' on dark photon dark matter by requiring that the IGM is not over-heated in the redshift range $z \in [6,10]$ (which would also assume that the temperature evolution of the IGM is well understood, which it is not \cite{Gaikwad:2020art}). We take the requirement $\varepsilon_{\rm inj}<1$ eV/ baryon.
We adopt a recent modeling of the mean free electron fraction from Ref.~\cite{Naidu:2019gvi}, which is consistent with high$-z$ quasar measurements. We plot the bounds obtained for both the homogeneous and strongly inhomogeneous scenarios in \Fig{fig:Rebnd}. The bounds obtained for $m_{A'}\sim10^{-13} - \few\tenx{-12}\ev$ are similar, but for smaller masses $m_{A'} \lesssim 10^{-14}\ev$ the difference in these treatments can be significant. 

We conclude this section by emphasizing that a detailed treatment of reionization will be needed to move beyond these two extremal examples, and will likely require numerical simulations to understand the correlations between ionized regions and over-densities.

\subsection{Star formation}

Dark-photon-induced heating may have strong implications for star formation. Since observations of high-redshift star formation do not yet exist, we simply outline here how the star formation rate might be suppressed, and the interesting regions of parameter space which might be testable using observations from 21cm telescopes or the James Webb telescope~\cite{gardner2006james,stiavelli2009first} in the near future.

Star formation is typically expected to be efficient when the kinetic temperature of the gas in a gravitationally bound object exceeds some threshold at which the cooling, and subsequent collapse, of gas becomes efficient. Assuming the kinetic temperature in the halo is approximately given by the virial temperature $T_k \simeq T_{\rm vir}$, one can relate this threshold directly to the host halo mass via~\cite{Shapiro:1998zp,Mena:2019nhm}
\begin{equation}
	T_{\rm vir}(M_h, z) \simeq 4.8 \times 10^4 \, {\rm K} \, \left(\frac{\mu}{1.22}\right) \left(\frac{M_h}{10^8 \, M_\odot \, h^{-1}} \right)^{2/3} \, \left(\frac{\Omega_m \, \Delta_c(z)}{\Omega_{m,z} \, 18\pi^2} \right)^{1/3} \, \left( \frac{1+z}{10}\right) \, .
\end{equation}
We have assumed the neutral hydrogen fraction is given by the IGM value, the halo profile is approximately given by that of a truncated isothermal sphere, and we recall that $\mu$ is the mean molecular weight ($\sim 1.22$ for a neutral medium).  The virial overdensity based on spherical collapse $\Delta_c(z)$ is given by~\cite{Bryan:1997dn}
\begin{equation}
	\Delta_c(z) = 18\pi^2 + 82 (\Omega_{m,z} - 1) - 39  (\Omega_{m,z} - 1)^2 \, ,
\end{equation}
where
\begin{equation}
	\Omega_{m,z} = \frac{\Omega_m (1+z)^3}{\Omega_m (1+z)^3 + \Omega_\Lambda} \, .
\end{equation}
Efficient star formation is only expected to proceed in halos with virial temperatures $\gtrsim 10^4$ K~\cite{Evrard:1990fu,blanchard1992origin,Tegmark:1996yt,Haiman:1999mn}, as this is the threshold for which molecular hydrogen cooling becomes efficient.  While there is uncertainty in both how to treat this threshold and where exactly it lies, it is clear that a sufficient number of collapsed objects with these masses must exist  in order to efficiently produce star forming regions.

If dark photon heating is active near the epoch of star formation, however, it may be possible that thermal pressure prevents the formation of star forming halos. That is to say, if the medium is heated to a sufficient degree, the Jeans mass may increase to a level where star forming halos would never have collapsed. To assess the extent to which this may occur, we compute the Jeans mass as a function of redshift and overdensity via 
\begin{equation}
	M_J(z, \Delta_b) = \frac{4\pi}{3}\bar{\rho}(z, \delta_b) \left(\frac{\lambda_J(z,\Delta_b)}{2}\right)^3 \, ,
\end{equation}
where  $\bar{\rho}$ is the average density in the radius of interest, and the Jeans length $\lambda_J$ is as defined in \Sec{sec:inhomoU}. As before, we will assume that the baryon density approximately follows that of the dark matter, and write $\bar{\rho} = \bar{\rho}_{b} \Delta_b (1 + \Omega_{\rm CDM} / \Omega_b) $.  The temperature evolution of a particular overdensity in the presence of dark photon heating is obtained by solving equations for the evolution of the matter temperature as in \Sec{sec:21cm}; thus, given a large-scale overdensity $\Delta_b$, one can determine the redshift-dependent evolution of the Jeans mass for any dark photon candidate. 

\begin{figure*}
	\includegraphics[width=0.495\textwidth]{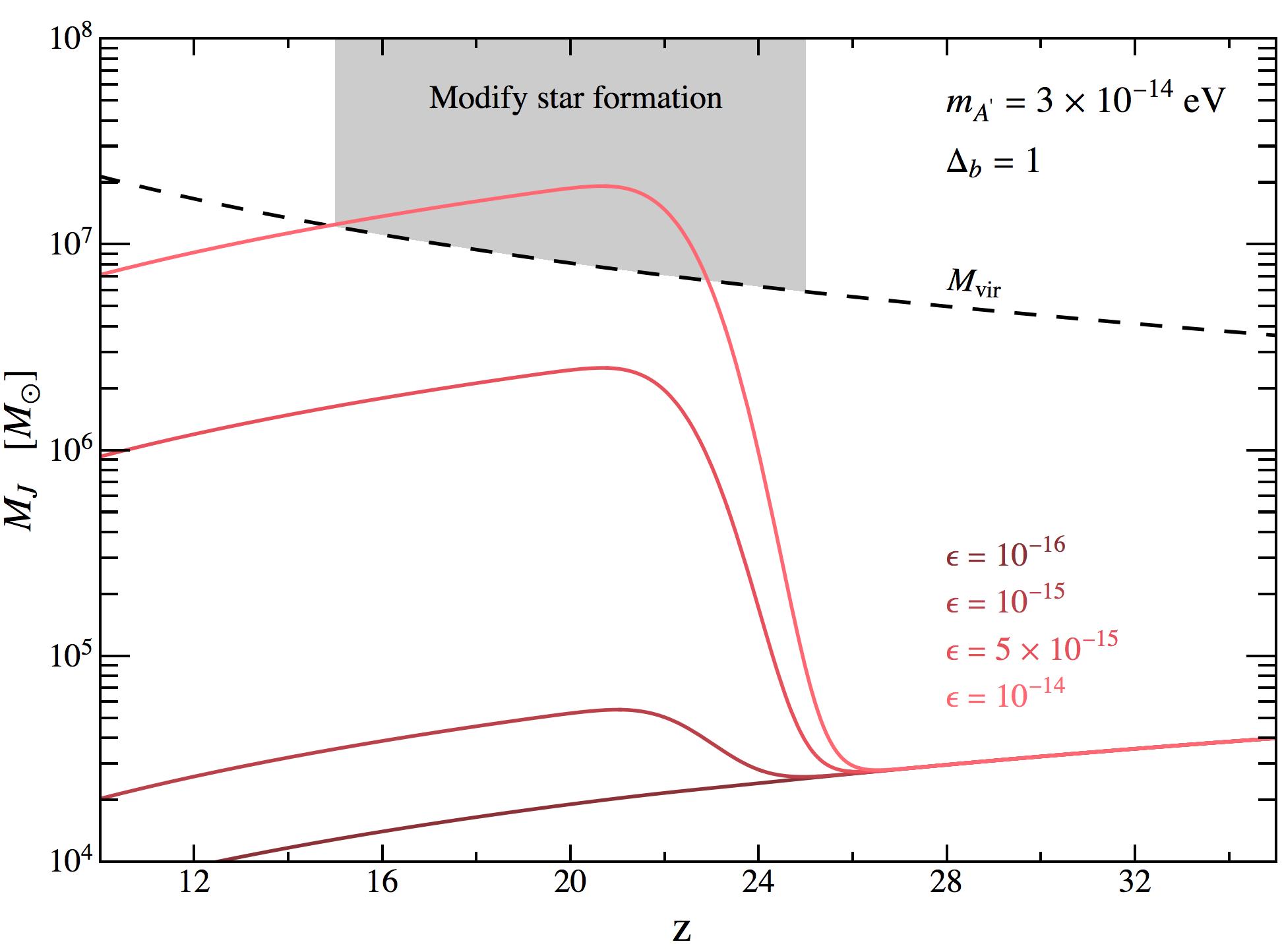}
	\includegraphics[width=0.495\textwidth]{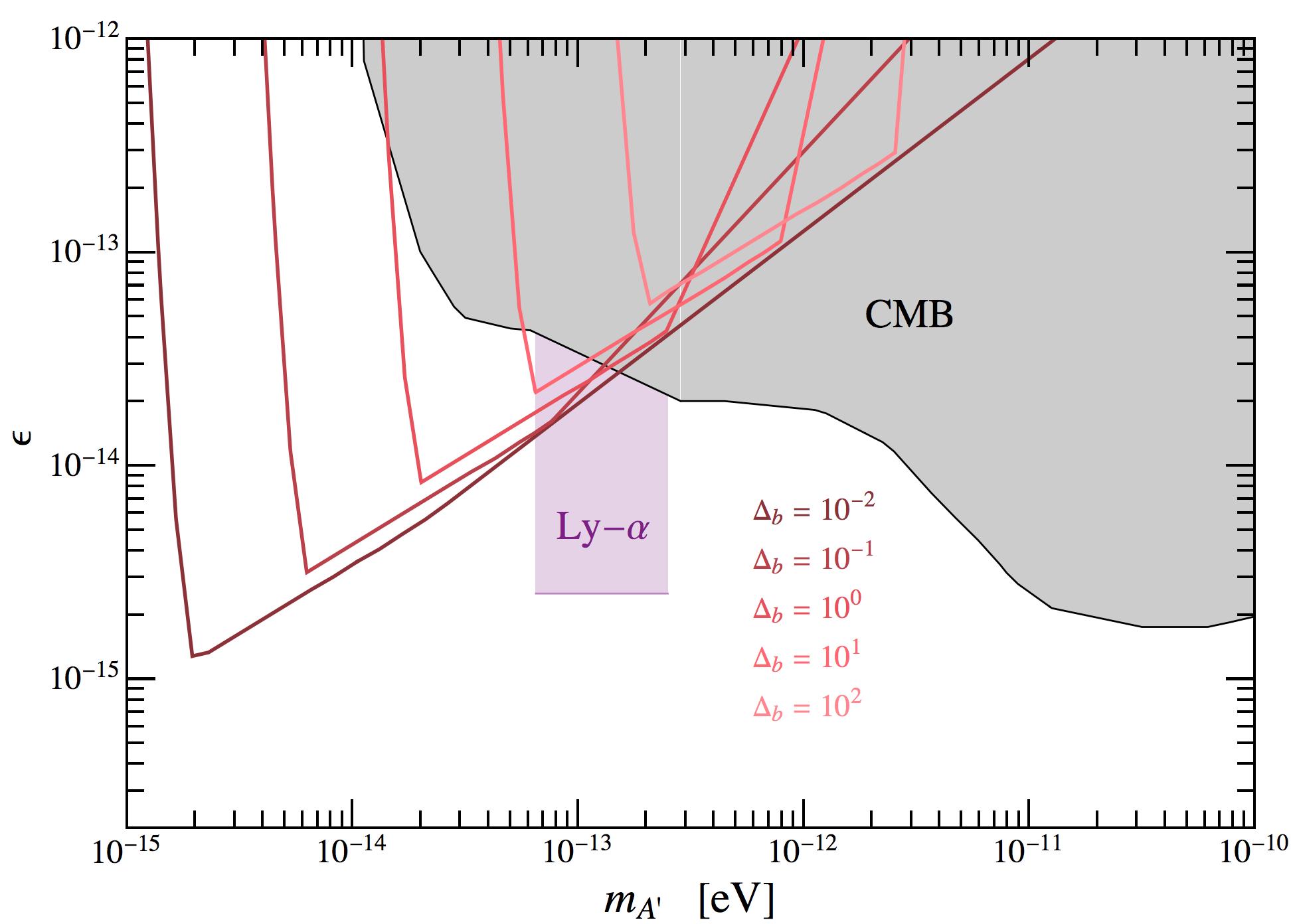}
	\caption{\label{fig:MJ} Left: Comparison of Jeans mass as a function of redshift assuming energy injected at $\Delta_b = 1$ from a dark photon with mass $3 \times 10^{-14}$ eV. The minimum virial mass required for star formation is shown in black dashed line (assuming star formation onsets when $H_2$ cooling becomes efficient, \ie when the virial temperature of a halo is $10^4$K). Should $M_J > M_{\rm vir}$ over the entire interval $15 \lesssim z \lesssim 25$, we expect star formation can be notably modified.  Right: Estimated sensitivity to the modification of the Jeans mass assuming modification to star formation is observed in various over and under-densities $\Delta_b$.}
\end{figure*}

We compare in the left panel of \Fig{fig:MJ} the evolution of the Jeans mass in an overdensity $\Delta_b = 1$ as a function of redshift, assuming a dark photon mass $3\times 10^{-14}$ eV and various mixings. We compare this mass scale with the redshift-dependent virial mass $M_{\rm vir}$, which is obtained by assuming the matter profile is given by a truncated isothermal profile with a temperature of $10^4$ K. Typically, star formation is expected to begin near redshifts $15 \lesssim z \lesssim 25 $, and thus if the Jeans mass sufficiently exceeds $M_{\rm vir}$ over this interval, star formation rates can be dramatically altered.  In order to assess the potential impact of dark photons on the star formation rate, we highlight in the right panel of \Fig{fig:MJ} the dark photon parameter space capable of suppressing star formation, which we define here as those models for which $M_J > M_{\rm min}^{\rm vir}$ over the entirety of the interval $z \in [15, 25]$, assuming star formation is observed in an isolated over-density $\Delta_b$\footnote{Stars in regions with $\Delta_b \ll 1$ at these redshifts should be extremely rare, and thus we expect the curves near $\Delta_b \gtrsim 1$ to be the most relevant to future surveys.}. This final assumption will likely not be met in most experiments, and in reality one should expect the sensitivity to be something of a $\Delta_b$-weighted average over these curves. However, if high-redshift 21cm experiments in the distant future find themselves capable of achieving 21cm tomography~\cite{Madau:1996cs}, perhaps the environmental dependence can be isolated. We also emphasize that this could be a detectable signature in the James Webb Space Telescope (JWST) which is likely to come online in the near future~\cite{gardner2006james,stiavelli2009first}.

\subsection{Dark Ages Energy Injection}\label{subsec:DA}

It was shown in Ref.~\cite{McDermott:2019lch} that strong constraints can be derived on dark photons that resonantly convert during the dark ages. Dark photons converting during this epoch will efficiently deposit their energy in baryons, heating the medium above the collisional ionization threshold and subsequently causing the medium to prematurely re-ionize. In the case of a homogeneous Universe, the energy injection is not sustained (\ie it occurs over a short period of time, after which atoms are allowed to cool), and some atoms will recombine. However, the asymptotic free electron fraction  will be significantly larger than in the case of $\Lambda$CDM. This can be strongly constrained using the CMB since the optical depth is extremely sensitive to the free electron fraction between recombination and reionization. 
This bound was recently criticized \cite{Caputo:2020bdy}, but we argue here that such concerns are unwarranted. 

\begin{figure*}
	\includegraphics[width=0.6\textwidth]{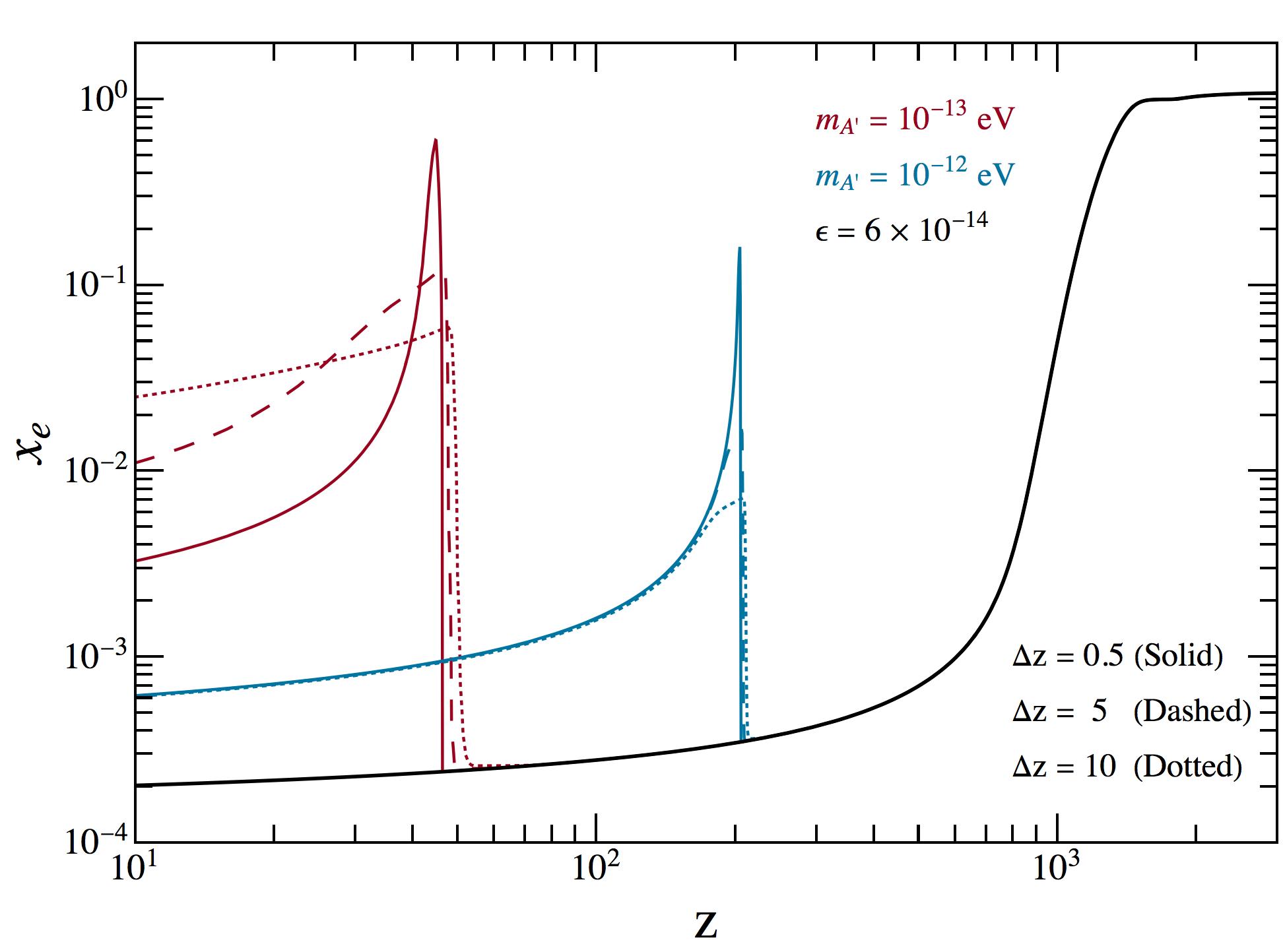}
	\caption{\label{fig:xeDA} Change in $x_e$ assuming energy is injected following a Gaussian distribution in $z$ with $\Delta z = 0.5$, 5, or 10, and for two masses and choices of mixing. Computation includes collisional ionization as described in Ref.~\cite{McDermott:2019lch}.}
\end{figure*}

There are two potential causes for concern. The first arises from the fact that the energy injection in Ref.~\cite{McDermott:2019lch} was treated assuming the redshift dependence followed a Gaussian distribution with $\Delta z = 0.5$. From \Fig{fig:den_violin} and \Fig{fig:fracE_violin} it is clear that the presence of inhomogeneities broadens the energy injection such that it spans $ \Delta z \sim 10$ at $z \sim 30$. This concern is easy to address.  In \Fig{fig:xeDA} we plot the evolution of the free electron fraction for two different dark photon parameters, assuming the energy injection can be modeled with a Gaussian distribution in redshift with $\Delta z = 0.5$ (solid),  $5$ (dashed), or $10$ (dotted). For the high redshift resonance  (\ie large $m_{A'}$) there is nearly no difference between all of the curves. At low redshifts, the $\Delta z = 10$ energy injection actually produces larger asymptotic values of the free electron fraction. The difference between these two sets of curves partially arises from the fact that a value of $\Delta z = 10$ is large relative to the resonance redshift itself, and partially because the efficiency of cooling processes depends on the maximal level of ionization obtained during the energy injection process (notice that the peak ionization fraction is larger for the narrow Gaussian). Since the energy injection is extremely sensitive to the value of $\epsilon$, we expect the differences here to produce tiny changes in the derived limit.

 The second concern is related to potential back-reaction. In the case of a homogeneous Universe it was shown that there can be no back-reaction which would cause the resonance to prematurely end~\cite{McDermott:2019lch}. The reason is simply that the resonant timescale is short relative to that of collisional ionization. The case of an inhomogeneous Universe is slightly more complicated, because the resonance timescale is not dictated solely by the width of the resonance, but rather by the rate at which $\Delta_{{\rm res}}$ sweeps through the overdensity PDF. However, in order for back-reaction to occur in the inhomogeneous scenario, not only must reionization in some over-density take place on short time scales, but free electrons produced in this process must quickly and efficiently diffuse to larger over-densities. Since the electrons are always non-relativistic, we expect this to be slow relative to the Hubble rate, and therefore we expect back-reaction to be negligible. A proper treatment of this process is beyond the scope of this work, and we leave a detailed exploration of such effects to future studies.

\subsection{Late-time Spectral Distortions}
 It was also shown in  Ref.~\cite{McDermott:2019lch} that strong bounds can be placed on dark photon dark matter resonantly converting into photons prior to recombination from the non-observation of spectral distortions in the CMB.  Should the gas be heated to a sufficiently high level, $y-$type distortions can still be induced after recombination; consequently, it is in general also possible to constrain dark photons from late-time heating of the IGM. 

\begin{figure*}
	\includegraphics[width=0.6\textwidth]{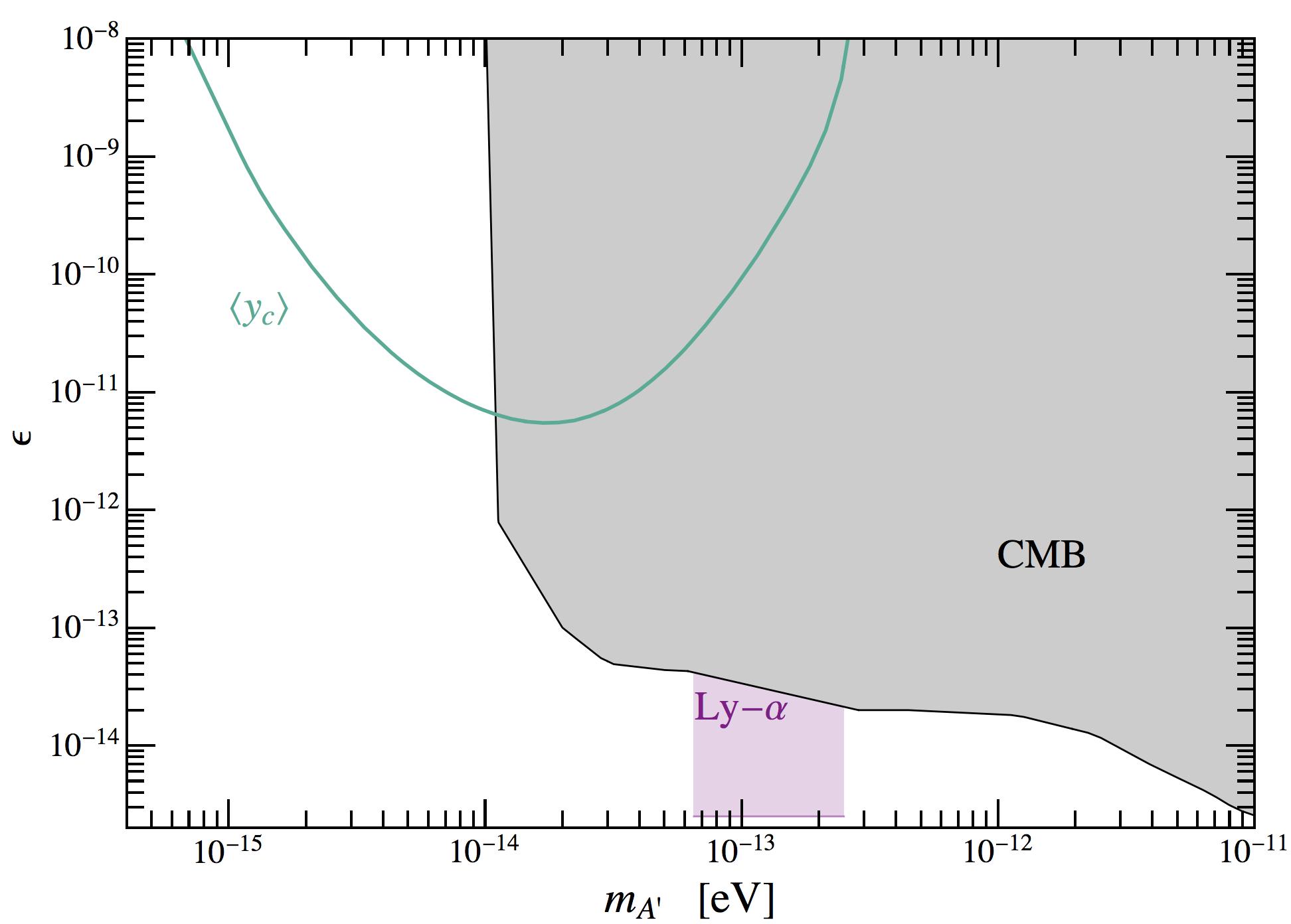}
	\caption{\label{fig:yCompton} Projected sensitivity for an experiment like PIXIE or PRISIM capable of measuring spectral distortions to a level of $y \sim 10^{-7}$.  Constraint is derived assuming reionization has not yet begun by $z \sim 10$, and dark photons supply the only source of heating. }
\end{figure*}

For heating that occurs at redshifts $z_{\rm reion} \lesssim z \lesssim z_{\rm recombination}$, one would naturally expect constraints from early reionization (discussed in \Sec{subsec:DA}) to be much stronger than those arising from late-time spectral distortions. Similarly, one must inject heat into the IGM at a level much greater than $\sim 1$ eV / baryon for redshifts $3 \lesssim z \lesssim 6$ in order to produce observable spectral distortions, implying Ly-$\alpha$ constraints will likely be stronger in this regime. At lower redshifts, one suffers from a vast array of astrophysical and cosmological uncertainty. In addition, the IGM is much hotter, meaning more energy must be injected in order to produce dramatic changes. Perhaps the most promising regime in which to look for these effects is at redshifts near reionization. In $\Lambda$CDM, reionization is already expected to contribute to $y$-type spectral distortions at the level of $\sim 10^{-7}$~\cite{Sunyaev:2013aoa}, which is slightly above the expected threshold for a futuristic CMB spectral distortion experiment like PIXIE~\cite{Kogut:2011xw} or PRISIM~\cite{andre2014prism}. In order to heuristically estimate what type of sensitivity these experiments might have to exotic heating from dark photon conversion, we evolve the temperature of the IGM as described in \Sec{sec:21cm}, assuming reionization begins only at $z < 10$ (note this is probably a rather optimistic assumption, but is sufficient for illustrative purposes).  We then compute the contribution of this heating to the Compton $y$ parameter~\cite{Sunyaev:2013aoa}, which when averaged over $\Delta$ yields
\begin{equation}
	\left< y_c\right> = \int_{z_{\rm min}}^{z_{\rm max}} dz \, \int d \Delta_b \, P_\Delta(z, \Delta_b) \, \frac{\sigma_T}{m_e} \frac{n^{\rm homo}_e(z) \, \Delta_b \, (T_k(\Delta_b) - T_{\rm cmb})}{H(z) \, (1+z)} \, .
\end{equation}
 We define this signal to be observable  when $\left< y_c\right> \geq 5 \times 10^{-7}$, although this number should be understood purely as a rough estimate, as uncertainties associated with reionization may alter the $\Lambda$CDM expectation by a factor of a few (exploring the detailed degeneracies between reionization histories and exotic energy injection from dark photons is beyond the scope of the current work). We show the sensitivity for a future PIXIE/PRISIM-like experiment in \Fig{fig:yCompton}. Depending on the details of reionization, this method may provide a probe of dark photons at masses below $\sim 10^{-14}$ eV. 

\section{Conclusion}\label{sec:conclusion}
In this work we have investigated the extent to which inhomogeneous structure can alter the cosmology of an ultra-light dark photon dark matter that injects energy after the formation of the CMB. This generalizes and expands upon existing work~\cite{McDermott:2019lch}, revisiting the robustness of existing bounds and investigating novel signatures which may be particularly sensitive to the inhomogeneous features arising at $z \lesssim 50$. Accounting for inhomogeneities has important implications for dark photons capable of resonantly converting to visible photons, since the resonance occurs when the dark photon mass equals the local plasma frequency of the ambient medium. Consequently, the homogeneous Universe approximation assumes (up to effects associated with reionization) a one-to-one mapping between the redshift of the resonance and the dark photon mass, while inhomogeneous structure at \eg $z\sim 5$ may produce resonant conditions for dark photons with masses spanning multiple orders of magnitude. The implications are most prominent for dark photons with lower masses, as such particles may appear to have no chance to undergo resonance in the homogeneous approximation, but in reality can resonantly convert in (and thus inject energy into) cosmic voids.

Using current observations of the Ly-$\alpha$ forest, we derived constraints in \Sec{sec:lymanA} on the excess heating of the IGM, extending on the work of Ref.~\cite{McDermott:2019lch} which had neglected the presence of inhomogeneities. Accounting for the presence of structure allows us to extend these constraints to masses approximately an order of magnitude smaller than were obtained in the homogeneous approximation, with only minor (but real) corrections to the constraining power obtained in the homogeneous limit. We found that the ability to constrain larger masses, however, was severely limited by the fact that quasar fluxes are preferentially absorbed in overdensities. 

In \Sec{sec:21cm} we investigated the extent to which future radio observations of the 21cm absorption signal at high redshift might be able to constrain resonant dark photon conversion. By conservatively assuming that future experiments observe the signal in absorption, which is a generic expectation for $z \gtrsim 12$, we show that experiments might be able to constrain mixings at the level of $\sim 10^{-15}$ for $m_{A'} \sim 10^{-14}$ eV. We expect that radio interferometers looking to measure the 21cm power spectrum will significantly improve the projected sensitivity of 21cm observations to light dark photons. Due to the complexity of modeling the contribution of inhomogeneous dark photon heating to the 21cm power spectrum, we defer a complete treatment of observables from this epoch to future work. 

Lastly, we have commented on other signatures that may arise at late times. In particular, we have discussed the epoch of reionization and the complications associated with understanding the distribution and evolution of plasma frequencies. In this case,  we have shown that  future data should allow one to study exotic energy injection during reionization, but robust analysis will require a detailed understanding of spatial inhomogeneities of ionized and non-ionized regions. We have also commented on the extent to which localized heating produced from dark photon absorption can increase the Jeans mass -- we have shown that efficient heating may actually raise the Jeans mass above the star forming threshold, particularly in under-densities, significantly suppressing star formation rates in these regions. Such effects may potentially be observable with the James Webb telescope or using  21cm tomography. Finally, we have shown that future experiments like PIXIE or PRISIM (which hope to measure spectral distortions in the CMB) may be sensitive to the heating of the IGM at redshifts just prior to reionization, potentially gaining unprecedented sensitivity to dark photons with masses $\sim 10^{-15}$ eV. 

In summary, properly accounting for the existence of baryonic structure at $z \lesssim 50$ is crucial for understanding the extent to which cosmology can probe the existence of ultra-light dark photon dark matter. The predominant effect of including these inhomogeneities is that it makes cosmological observations sensitive to a wider range of masses,  allowing to probe lower masses, while relaxing some constraints at larger mass. Furthermore, we emphasize that experiments which themselves are inherently sensitive to the inhomogeneous state of the Universe could provide striking signatures that are unique to light dark photon dark matter, and thus offer a tantalizing opportunity for a positive detection, should they exist.

 \acknowledgments{We thank Nick Gnedin, Hongwan Liu, Josh Ruderman, Andrea Caputo, and Siddarth Mishra-Sharma for discussions. SJW thanks Luisa Lucie-Smith and Andreu Font-Ribera for their valuable discussions on structure formation and the Lyman$-\alpha$ forest, respectively. SJW and SR acknowledge support from the European Union's Horizon 2020 research and innovation program under the Marie Sk\l{}odowska-Curie grant agreements No.\ 690575 and 674896. SJW acknowledges support under Spanish grants FPA2014-57816-P and FPA2017-85985-P of the MINECO and PROMETEO II/2014/050 of the Generalitat Valenciana. SR also acknowledges support from the ``Spanish Agencia Estatal de Investigaci\'on'' (AEI) and the EU ``Fondo Europeo de Desarrollo Regional'' (FEDER) through the project FPA2016-78645-P; and through the Centro de Excelencia Severo Ochoa Program under grant SEV-2016-0597. Fermilab is operated by Fermi Research Alliance, LLC under Contract No.~DE-AC02-07CH11359 with the United States Department of Energy.}

\bibliography{biblio}
\bibliographystyle{JHEP}

\appendix

\section{Comparison to Previous Work}\label{sec:comparison}
While this work was being completed, Ref.~\cite{Caputo:2020bdy} appeared on the arXiv, presenting ideas that overlap with some of those discussed in \Sec{sec:lymanA}. Specifically, Ref.~\cite{Caputo:2020bdy} writes down a formalism to account for the injection of energy from dark photon dark matter resonantly converting to photons in the presence of inhomogeneities, and uses this formalism to derive constraints from Ly-$\alpha$ observations of the IGM temperature (as done in \Sec{sec:lymanA}). Our results, however, are not in exact agreement\footnote{We thank the authors of \cite{Caputo:2020bdy} for discussions on these topics.}.
	
One of the key points lying at the heart of this disagreement is the understanding of the final fate of the injected energy. In this paper, we have worked under the assumption that energy injection is a local phenomenon. That is to say, energy injection at a given $\Delta_b(\vec{x})$ does not have any significant impact on a different $\Delta_b(\vec{x}^\prime)$. The formalism of Ref.~\cite{Caputo:2020bdy} on the other hand implicitly assumes that energy injected anywhere is quickly thermalized everywhere. If the assumption of Ref.~\cite{Caputo:2020bdy} were correct, there would be two immediate implications: $\emph{(i)}$ when computing the energy injection per unit baryon, one should neglect the $\Delta_b$ dependence in the baryon number density, since energy is ultimately shared between under- and over-densities (this directly modifies the importance of under- and over-densities in the calculation of $\varepsilon_{\rm inj}$), and $\emph{(ii)}$ one should neglect the observational sensitivity to $\Delta_b$ (\eg one does not need to be concerned with the optical depth of the Ly-$\alpha$ photons, as discussed in \Sec{sec:lymanA}). We believe, however, that this cannot be the case, as the photons produced in resonant transitions have short mean free paths~\cite{McDermott:2019lch}, meaning they are instantaneously absorbed by the local medium, and the electrons that absorb the energy remain non-relativistic. In order for efficient thermalization across inhomogeneities to occur, the diffusion timescale must be small relative to Hubble time, which seems inherently at odds with the fact that electrons are non-relativistic. 
 
\begin{figure*}
	\includegraphics[width=0.55\textwidth]{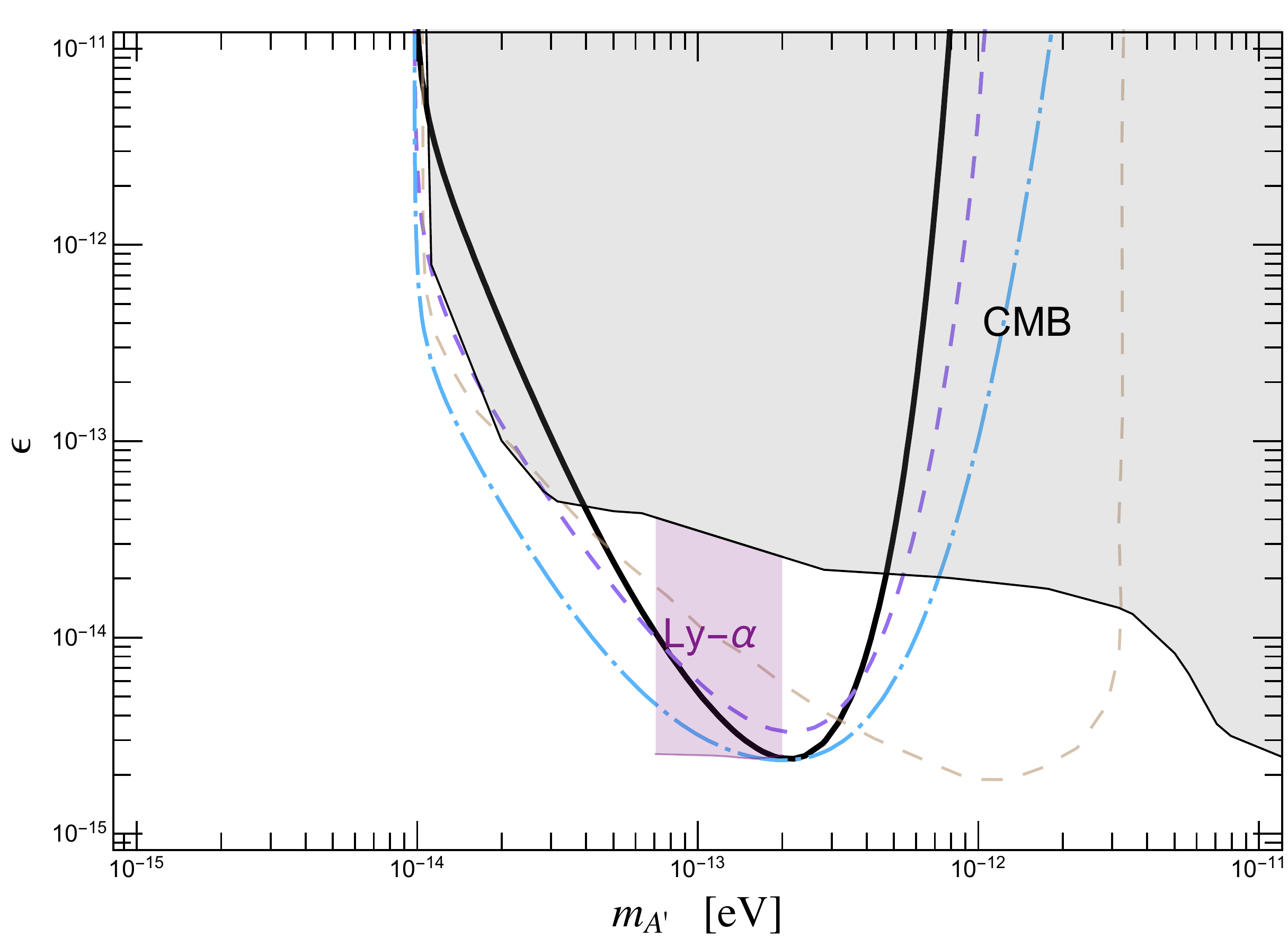}
	\caption{\label{fig:lyBND_Comp} Our result from \Fig{fig:lyBND} obtained with \Eq{eq:S2} using $T_k = 10^4$ K (black solid), or using the adiabatic temperature of the IGM (neglecting the heating induced during reionization, dashed purple). The result is also compared with the bound that would be obtained if we neglect the sensitivity function, \ie $\mathcal{S}(\Delta_b, z) = 1$ (blue dot-dashed). A comparison is made to the constraint obtained by~\cite{Caputo:2020bdy} (light brown, dashed). }
	\end{figure*}

 In addition to the apparent disagreement about the potential thermalization of the injected energy, Ref.~\cite{Caputo:2020bdy} uses the baryonic power spectrum of hydrodynamic simulations to determine the Jeans scale and baryon PDF. The value of the Jeans scale is determined to be around $\sim 10$ kpc at low redshifts. The value found using \Eq{eq:Jeans} is nearly two orders of magnitude larger at $z \sim 6$, which has large implications for the constraints that can be derived using the Ly-$\alpha$ forest. In \Fig{fig:lyBND_Comp} we show our result from \Fig{fig:lyBND} in the thick black line; we contrast this with the impact of using  \Eq{eq:Jeans} with the adiabatic temperature of the IGM (neglecting heating from reionization, purple dashed). We also show the implications of neglecting the sensitivity function (\ie taking $\mathcal{S}(\Delta_b, z ) = 1$) (blue, dot dashed). We compare to the bound derived in Ref.~\cite{Caputo:2020bdy}. We find strong differences in the constraints derived at low masses when varying the Jeans scale between $10$ kpc and $\sim 1$ Mpc, and strong differences at high masses when the effect of the Ly-$\alpha$ sensitivity is not included.

\end{document}